\documentclass[preprint,authoryear,12pt]{elsarticle}

\usepackage{amssymb,amsthm,amsmath,amsbsy,epsfig,float,natbib,lscape,makeidx}
\usepackage{multirow}
\usepackage{color,algorithm}
\definecolor{MyDarkBlue}{rgb}{0,0.08,0.45}
\definecolor{yellow}{rgb}{0.99,0.99,0.70}
\definecolor{white}{rgb}{1.0,1.0,1.0}
\definecolor{black}{rgb}{0.00,0.00,0.00}
\definecolor{grey1}{rgb}{0.9,0.9,0.9}
\definecolor{grey2}{rgb}{0.8,0.8,0.8}

\def\argmin{\mathop{\rm argmin}}
\def\arginf{\mathop{\rm arginf}}

\newcommand{\real}{\ensuremath{\mathbb{R}}}

\def\argmin{\mathop{\rm argmin}}

\def \P {\mathcal{P}}

\newtheorem{algo}{Algorithm}

\newtheorem{remark}{Remark}

\begin{document}
	\begin{frontmatter}
		\title{Discovering Common Change-Point Patterns in Functional Connectivity Across Subjects
			\footnote{This paper was presented in part at the conference on Information Processing in Medical Imaging (IPMI), 2017
		}}
		
		\author[FSU]{Mengyu Dai}
		\author[Rochester]{Zhengwu Zhang}
		\author[FSU]{and Anuj Srivastava}
		
		\address[FSU]{Department of Statistics, Florida State University, Tallahassee, FL, USA}
		\address[Rochester]{Department of Biostatistics and Computational Biology, University of Rochester Medical Center, Rochester, NY}
		
		\begin{abstract}
			This paper studies change-points in human brain functional connectivity (FC) and seeks patterns that are common across multiple subjects under identical external stimulus. FC relates to the similarity of fMRI responses across different brain regions
			when the brain is simply resting or performing a task. While the dynamic nature of FC is well accepted, this paper develops a formal statistical test for finding {\it change-points} in 
			times series associated with FC. It represents short-term connectivity by a symmetric positive-definite matrix, and uses a Riemannian metric on this space to develop a graphical method for detecting change-points in a time series of such matrices. It also provides a graphical representation of estimated FC for stationary subintervals in between the detected change-points. Furthermore, it uses a temporal alignment of the test statistic, viewed as a real-valued function over time, to remove inter-subject variability and to discover common change-point patterns across subjects. This method is illustrated using data from Human Connectome Project (HCP) database for multiple subjects and tasks. 
		\end{abstract}
		
	\end{frontmatter}
	
	\tableofcontents{}
	
	\section{Introduction} 
	Learning about structural and functional connectivity in human brain is of great interest from many perspectives. The recent Human Connectome 
	Project (HCP)  investigates these connectivities in order to understand brain functionality and  its relation to the cognitive abilities of individual subjects. Functional connectivity (FC) is observed using functional MRI (fMRI) data and is believed to capture interactions  between brain regions with or without external stimuli. It is different from structural connectivity
which describes  anatomical links between brain regions and is estimated using structural MRI or diffusion MRI. FC is defined  as a set of {\it statistical dependencies among remote anatomical regions due to neurophysiological events} (\cite{friston-bc:2011}).
	These dependencies are expressed as quantifications of similarity, or {\it correlations},  between simultaneous functional measurements
	of neuronal activities across regions in the human brain.  Functional MRI measures the 
	blood oxygen level dependent (BOLD) contrast signals of each brain voxel over a period of time.
	The short-term FC is often represented as a covariance or correlation 
	matrix of fMRI data over a small time window, with the matrix size equal to the number of brain regions being considered. 
	In the early days,  FC associated with individual tasks or stimuli was treated as fixed or static over observation time. 
	However, recent studies (\cite{Hutchison-etal:2013,Monti2014427,Hindriks2016242}) have revealed strong evidence 
	that FC is a dynamic process and evolves over time, even in the resting state.  
	Our primary interest is to investigate and characterize the dynamic nature of 
	FC between different anatomical regions during performances of certain tasks and in resting states.
	A dynamic FC model is bound to provide broader insight into fundamental mechanism of brain networks (\cite{Lindquist2014531}).
	For a review on recent progress in analyzing FC, please refer to \cite{friston-bc:2011,Hutchison-etal:2013}.

	A critical part in modeling dynamic FC is to quantify changes that occur over time during observation intervals. 
	For example, some studies have developed tests for temporal changes in coherence pattern of multiple regions of 
	interest (ROIs) \cite{Poldrack01032007}.  
	\cite{CRIBBEN2012907}  introduced Dynamic Connectivity Regression (DCR), which 
	is a data-driven technique for detecting temporal change 
	points in functional connectivity between brain regions.
	Later on, \cite{fncom.2013.00143} presented a modified DCR on single-subject data that increases accuracy 
	with a small number of observations and reduces the number of false positives. In case FC is represented by a covariance or a correlation matrix of fMRI signals, 
	a sliding (time) window is commonly used to estimate the dynamic FC as a time series of such matrices. 
	\cite{Lindquist2014531} discuss some alternatives to moving window-based correlation estimation. \cite{fnins.2015.00285} have proposed an advancement of DCR into a dynamic connectivity detection algorithm that uses statistical models at node level and can handle much larger number of regions than DCR. More recently, \cite{PMID:26739105} proposed a method for detecting change-points in correlation networks that requires minimal distributional assumptions. \cite{JEONG2016353} presented a semiparametric mixture model which is less sensitive to the window size selection in detecting change-points in large-size functional networks of resting-state fMRI. \cite{10.1007/978-3-319-59575-7_28} studied the problem of change-point detection based on Bayesian inference and genetic algorithms.

	While several of these studies focus on testing whether FC is static or dynamic, or seek temporal change-points for  {\it individual} subjects, we additionally analyze FC in a {\it group} of individuals and 
	seek some common patterns of FC during performances of identical tasks. We take different partitionings of human brain into functional units or regions of interests (ROIs) and represent short-term FC as symmetric, positive definite matrices (SPDMs), with entries denoting covariances of localized BOLD signals. 
	We use a sliding window technique to segment multivariate time series into overlapping blocks, and compute a covariance matrix for each block. 
	The original time series data can thus be converted into a time-indexed sequence of SPDMs and one of the goals is now to detect if these 
	sequences denote a stationary process or if the underlying 
	probability distribution switches at one or multiple time points. 
	The actual test is performed using a nonparametric graphical approach, 
	introduced in \cite{chen2015}, that uses minimal 
	spanning trees (MSTs) connecting the observed SPDMs in a certain metric space. Even though this approach is derived for $i.i.d.$ samples, we empirically demonstrate that it is also effective in case of dependent time series data. We utilize a Riemannian structure of the space of SPDMs to define and compute geodesic distances between covariance matrices, and to facilitate the construction of MSTs. Computation of edge length 
	distributions across temporal partitions of nodes in  MSTs lead to a test statistic for change-point detection. 
	The final algorithm depends on a number of parameter choices for forming and testing covariance trajectories. 
	While the proposed method is naturally dependent on the choices of these parameters, we demonstrate that 
	the results are relatively stable with respect to these choices as long as they are in a reasonable 
	range as dictated by the experimental setting. 
	
	The change-point patterns for a group of people are expected to be similar, especially when the fMRI data are recorded in a well-designed experiment
	where tasks are sequentially performed at pre-determined time blocks. However, different subjects may react differently to external stimuli and the 
	response times can be different. 
	Additionally, data variability and algorithmic imprecision add to the variability in the detection results. 
	Thus, it is naturally difficult to discern the commonality of change-point patterns across subjects from the 
	raw change-point inferences. One solution to reducing inter-subject variability in the time domain is to perform temporal alignment, to allow discoveries of underlying patterns.  We apply temporal alignment at the level of test statistics to detect and discover some common patterns of FC dynamics for a group of individuals. We further validate these patterns by comparing them with block designs governing the task-related external stimuli using the HCP data. The proposed methodology is also 
	comprehensively tested with simulated data and validated by comparisons with the ground truth. The novel contributions of this paper are as follows:
	\begin{enumerate}
		\item  It introduces an end-to-end pipeline for change-point detection in individual subject data, starting from raw fMRI data all the way to detected change-points in FC.
		
		\item It represents short-term connectivity using covariance matrices and long-term data as covariance trajectories, and utilizes the Riemannian geometry of SPDM space for developing test statistics.
		
		\item It adapts the graph-based change-point detection algorithm presented
		in \cite{chen2015} to this problem area, and demonstrates the feasibility of this solution on a large set of 
		real and simulated data, despite its theoretical limitations.
		
		\item It applies the proposed framework to the state-of-the-art HCP task data, and 
		discovers change-point patterns across subjects using temporal alignment of test statistics.
		
	\end{enumerate}

	The rest of this paper is organized as follows. In Section 2, we describe  mathematical components of the proposed pipeline, including our representation of the dynamic FC,  a Riemannian structure on the space of SPDMs and a graphical approach for detecting change-points. In Section 3, we introduce a temporal alignment method to discover common change-point patterns. Section 4 presents experimental results using both simulated and real fMRI data. Section 5 closes the paper with a short conclusion and discussion.
	
	\section{Methodological Pipeline}
\begin{figure} 
		\centering  
		\includegraphics[width=5.5in]{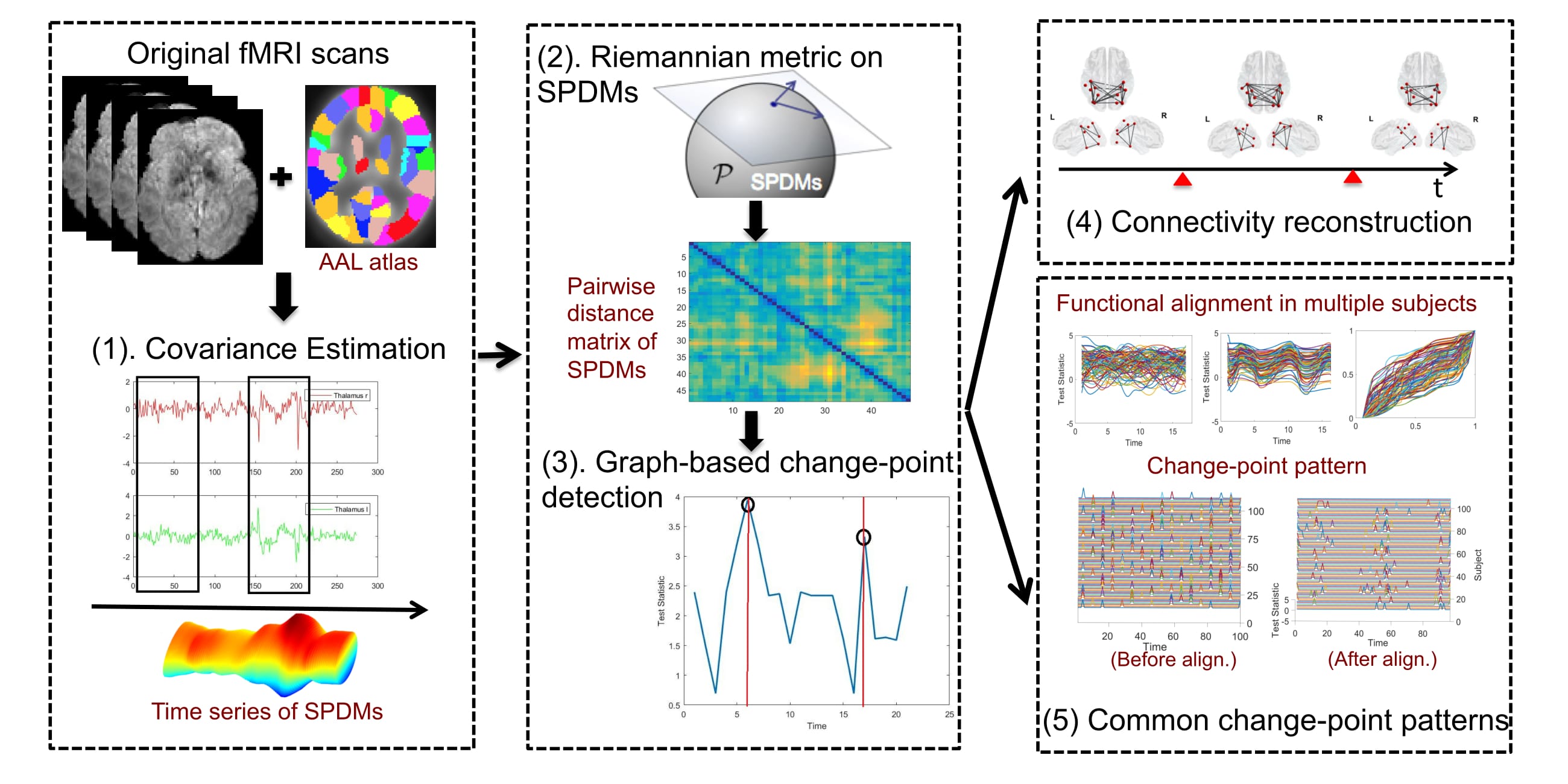}  
		\caption{A systematic overview of our proposed framework for FC change-point study.} \label{fig:overview}	\end{figure}

	The proposed pipeline, for discovering 
	dynamic patterns in FC over a population, is made up of several current and novel components. Figure~\ref{fig:overview} shows a systematic overview of the proposed 
	pipeline. We describe the individual pieces one by one in this section.
	
	\subsection{FC Estimation Using Covariance Matrices} 
	The first problem is to estimate a correlation or a covariance matrix that represents short-term FC between anatomical regions of interest. 
	To specify this matrix, we pick $n$ ROIs and 
	then represent the FC over a short period using a $n \times n$ covariance matrix of corresponding BOLD signals. 
	In this paper we use the approach from \cite{LEDOIT2004365} to estimate the covariance matrix. Here the estimator $\Sigma_*$ is an asymptotically optimal convex linear combination of the sample covariance matrix $S_*$ and the identity matrix $I$, i.e. $\Sigma_*=\rho_1I+\rho_2S_*$, and the optimal weights $\rho_1$ and $\rho_2$ are estimated from the data. This estimator is distribution-free and has a simple explicit formula that is easy to compute and interpret. The optimality is defined with respect to a quadratic loss function, valid when the number of observations and the number of variables go to infinity together. Extensive Monte Carlo studies
	confirms that the asymptotic results tend to hold well even for a reasonable finite sample. \cite{Brier2015} have applied this technique in an analysis of dynamic FC. There are other ways of estimating covariance matrix, see for example \cite{hinne-etal-PLOS:2015}, and we can easily substitute them 
	here as needed. 
	
	We segment the original multivariate time series into overlapping blocks of equal size. We will use $W$ as the window size for estimating covariance matrices and $S$ as the step size for sliding windows:
	$W$ should be large enough to result in a positive definite 
	covariance, but not too large to smooth over subtle changes that are critical to assessments of temporal evolution of FC. 
	Additionally, $S$ should be large enough to reduce dependency in successive covariance matrices for a better change-point detection. These quantities are determined in 
	practice using trial and error, especially with the help of simulated data since the ground truth is available 
	in that situation. This issue is discussed in more detail later
	in the paper.

	\subsection{Riemannian Structure on Symmetric Positive-Definite Matrices (SPDMs)}
	In order to quantify differences in FC, represented by the corresponding covariance matrices, 
	we need a metric structure on the manifold of covariance matrices or SPDMs.  While there are several Riemannian structures used in the literature (including one in \cite{Pennec:2006}), we briefly summarize the idea used in this paper. More details can be found in \cite{su-etal-JIVC:2012} and \cite{zhang-etal:2015}.
	
	Let ${\mathcal{P}}$ be the space of $n \times n$ SPDMs, and let $\tilde{\cal P}$ be its subset of matrices with determinant one. In this approach one imposes 
	separate distances on the determinant one matrices and the determinants themselves.
	Recall that for any square matrix $G$ with unit determinant, we can write it as a product of two square matrices $G = PS$,  
	where $P \in \tilde{\cal P}$ and $S \in SO(n)$  ($SO(n)$ is the set of all rotation matrices). This is called the 
	{\it polar decomposition} and motivates the analysis of $P$ by representing it as a $G$ after removing $S$. More formally, one obtains a 
	distance on $\tilde{\cal P}$ by identifying it with the quotient space $SL(n)/SO(n)$. This identification is based on the map $\pi: SL(n)/SO(n) \to \tilde{\mathcal{P}}$, given by 
	$\pi([G]) = \sqrt{\tilde{G}\tilde{G}^t}$, for any $\tilde{G} \in [G]$.  The notation $[G]$ stands for all possible rotations of the matrix $G$, given by 
	$[G] = \{ GS| S \in SO(n)\}$. 
	The inverse map of 
	$\pi$ is given by: $\pi^{-1}(\tilde{P}) = [\tilde{P}] \equiv \{\tilde{P} S|S \in SO(n)\} \in SL(n)/SO(n)$. 
	Skipping further details, this process leads to the following geodesic distance between points in $\tilde{\mathcal{P}}$. 
	For any $\tilde{P}_1, \tilde{P}_2 \in \tilde{\mathcal{P}}$:
	\begin{equation} 
	d_{\tilde{\cal P}}(\tilde{P}_1, \tilde{P}_2) =  \|A_{12}\|, \ \ 
	\mbox{where}\ \  A_{12} = \log(P_{12}),\ \ \ P_{12} =\sqrt{\tilde{P}_1^{-1} \tilde{P}_2^{2} 
	\tilde{P}_1^{-1}}\ .
	\end{equation}
	Since for any ${P}\in {\mathcal{P}}$ we have $\det({P}) > 0$, we can express ${P}$ as the pair $(\tilde{P}, \frac{1}{n}\log(\det({P})))$ 
	with $\tilde{P} =\frac{{P}}{\det({P})^{1/n}}\ \in \tilde{\cal P}$. Thus, ${\mathcal{P}}$ is identified with the product space of 
	$\tilde{\cal P} \times \real$ and we take a weighted combinations of distances on these two components to reach a metric on $\mathcal{P}$:
	\begin{equation}
	d_{{\mathcal{P}}}(I, {P})^2 = d_{\tilde{\mathcal{P}}}(I, \tilde{P})^2 +\frac{1}{n}\ (\log(\det({P})))^2.
	\end{equation}	
	For any two arbitrary covariances ${P_1}$ and ${P_2}$, 
	let ${P_{12}} = {P_{1}^{-1}}{P_2}S_{12}$ for the optimal $S_{12} \in SO(n)$ 
	(using Procrustes alignment). For this
	${P_{12}} \in {\mathcal{P}}$, we have $\det({P_{12}}) = \det({P_2})/ \det({P_{1}})$ 
	and define $\tilde{P}_{12} = P_{12}/\det(P_{12})$.
	Therefore, the resulting squared geodesic distance between ${P_1}$ and ${P_{2}}$ is:  
	\begin{equation}
	d_{{\mathcal{P}}}({P_1}, 
	{P_2})^2 = d_{\tilde{\mathcal{P}}}(I, \tilde{P}_{12})^2 +\frac{1}{n}\ (\log(\det({P_2}))-\log(\det({P_1})))^2.
	\end{equation}
	Once we have a distance between covariances, we can use that to define and compute sample means on ${\cal P}$ as
	follows. The sample Karcher or Fr\'echet mean of a given set of covariances is defined to be: 
	\begin{equation}
	\bar{P}=\arg\min_{{P} \in {\cal P}} 
	\left(\sum_{i=1}^{n} d_{{\mathcal{P}}}({P},P_i)^2 \right),
	\end{equation}
    where $\{P_i \in {\cal P}\}$ is set of given covariances. 
	The algorithm for computing this mean is a standard one and is not 
	repeated here. We refer the reader to \cite{su-etal-JIVC:2012} for more details.

	\subsection{Change-Point Detection Using MST}
	The next issue is to detect change-points in a SPDM time series using a metric-based approach. We adapt a nonparametric graph-theoretical method introduced by \cite{chen2015} for solving this problem. There have been earlier works on graph-based methods for a two-sample test, but these authors extend the framework to arbitrary metric spaces. Even though this method is theoretically designed for $i.i.d.$ samples from their respective distributions, extensive experiments establish its usefulness even for dependent time series.
	
	Let $X_1, X_2,$ $\dots, X_L$ be a random, time-ordered sequence taking 
	values on a Riemannian manifold $M$, and let $F_0$ and $F_1$ denote two probability distribution  functions on $M$. We are interested
	in testing the null hypothesis that all $X_t$s are samples from the same distribution, against an alternative
	that all points before and after a certain time $\tau$ follow different distributions, respectively. That is, 
	$$
	H_0: X_t \sim F_0,\ t=1,2,\dots,L, \ \ \ \ \ \ \ \ \ \
	H_1: \exists \ 1\leq \tau <  L,\ \ X_t \sim \left\{ \begin{array}{cc} F_1, & t > \tau \\
	F_0, &\ \ \mbox{otherwise}\ , \end{array} \right. 
	$$
	This test is phrased for a single change-point at time $\tau$, but we will repeat it for different $\tau$s separated by 
	a certain minimum spacing
	to detect multiple change-points. 
	
	The test statistic $Z(\tau)$ for this test is computed as follows. 
	First compute all the pairwise distances between all $\{X_t\}$ under the chosen metric on $M$. 
	Then, use these distances to form a minimal spanning tree (MST) connecting the set $\{X_t\}$ in $M$. 
	An MST is a connected graph such that: (1) each $X_t$ is a node in the graph (i.e., connected to at least one 
	other point in the set,
	(2) if any two points are connected by an edge then the weight of that edge is 
	given by the pairwise distance between those points, and (3)  the sum of weights of all connected
	edges is the smallest amongst all such graphs. 
	Note that the MST is independent of the ordering of the given points, and can be computed/stored for the whole 
	sequence in one shot. 
	To test whether there is a change-point at $\tau \in \{1, 2, \dots, L-1\}$, 
	we divide $X_t$s into two groups: $\{X_1,\dots, X_{\tau}\}$, and $\{X_{\tau+1}, \dots, X_L\}$. 
	Let $R(\tau)$ represent the number of edges that connect 
	points across these two groups in MST. Intuitively, if $\tau$ is a change-point, with before and after samples 
	being homogeneous in their respective sets, then the two groups represent relatively disjoint
	clusters of points and only a small number of edges connect across those clusters. In contrast, if the two groups are
	from the same distribution, then we expect a large number of edges in MST going across the two groups. 
	\cite{chen2015} show that  under $H_0$, i.e. there is no change-point at $\tau$, the quantity $R(\tau)$ 
	has a distribution with mean  and variance given by:  
	$$E( R(\tau))= p_1(\tau)|G|,$$ $$Var( R(\tau))= p_2(\tau)|G|+ (\frac{1}{2} p_1(\tau)-p_2(\tau))\sum_t |G_t|^2 + (p_2(\tau)-{p_1}^2(\tau))|G|^2,$$ where $p_1(\tau)= \frac{2\tau(L-\tau)}{L(L-1)} $ and $p_2(\tau)=\frac{4\tau(\tau-1)(L-\tau)(L-\tau-1)}{L(L-1)(L-2)(L-3)}$. 
	Here, $|G|$ is the total number of edges in $G$ and 
	$|G_t|$ is defined as the number of edges in subgraph of $G$ containing all edges that connect to node $X_t$.
	Now, we can standardize $R(\tau)$ using $Z(\tau)=-{(R(\tau)- E(R(\tau)) \over \sqrt{Var(R(\tau))}}$ to reach the test statistic.
	This test statistic is then tested for significance at a given confidence 
	level, for accepting or rejecting the null hypothesis. 
	According to \cite{chen2015},  a  value of
	$Z(\tau) \geq 3$ implies a change-point detection at $0.95$ confidence level, and $Z(\tau) \geq 4$ implies a change-point 
	detection at $0.99$ confidence level. One can also use a permutation test to set a threshold for change-points at a certain significant level. 
	However, when using permutation tests on simulated data, we found that the resulting thresholds are actually quite similar to the values stated above. 
	Therefore,  we simply use the value of 3 for all experiments in order to save computational cost. The change-point detection process is summarized in Algorithm \ref{algo:change-point-detect}.

	\begin{remark}
		In this setup, the choice of parameters such as $W$, $S$, and $L$ plays an important role. 
		As mentioned previously, $W$ should be large enough to obtain a positive-definite matrix centered at each time, 
		and $S$ should be large enough to avoid the smoothing over of covariance variability. Similarly, $L$ should be large so we have sufficient sample size for change-point detection. On the other hand, it should not to be too large to contain 
		multiple change points in each estimation. In our implementations, we choose these parameters on the basis of 
		overall time length of the fMRI signal $(L_{total}$), number of ROIs ($n$), and these above considerations. 
		Furthermore, we study the effect of changes in $W$ on the detection results and demonstrate that the 
		results are quite stable with respect to this parameter. (One can do the same for $S$ also if the data is rich enough 
		to allow that study, but we expect a similar stability within reasonable limits). 
		So, we have generally used parameter values $W=16$ and $S=6$ in our experiments in which
		$L_{total} \approx 250$ and $n = 4$ to $15$.   
	\end{remark}

	\begin{algo}\label{algo:change-point-detect}
	{\bf Change-Point Detection for SPDM Time Series}
	
		\begin{enumerate}   
			
			\item For a given {\bf time ordered} set of points $\{ X_t \in M, t=\tau-0.5L, \tau-0.5L+1, \dots, \tau+0.5L-1\}$, compute $D$, the $L\times L$ pairwise geodesic distance matrix. Here $\tau$ takes value on the 
			set $\{0.5L+1, 0.5L+2, \dots, L_{total}-0.5L+1\}$ and $L_{total}$ is the length of the full covariance sequence.
			
			\item Use $D$ to form a minimal spanning tree (MST) connecting all the points. 
			
			\item Count $R(\tau)$, the number of edges in this MST between first group:  $\{X_{\tau-0.5L},\dots,X_{{\tau}}\}$, and the second
			group $\{X_{\tau + 1},\dots,X_{\tau+0.5L-1}\}$.  
			
			\item Calculate $E(R(\tau))$ and $Var(R(\tau))$, and use them to standardize $R(\tau)$,
			resulting in $Z(\tau)$.  At $0.95$ confidence level, use $Z(\tau)\geq3$ for detecting a 
			change point. By construction, the separation between any two change-points will be at least $W$ in
			this framework. 
			
			\item Set $\tau=\tau + 1$ and repeat Step $1$ - Step $4$ until $\tau=L_{total}-0.5L+1$.
			
			\begin{remark}
				Since $Z(\tau)$ can be potentially high for a number of neighboring 
				$\tau$s, we select the local maximizer of $Z(\tau)$ as the detected change-point.
				That is, we find time points where  -- function $Z(\tau)$ is increasing before it and decreasing after it.
			\end{remark}
			
		\end{enumerate}
	\end{algo}
	
\begin{figure} 
	\centering  
	\begin{tabular}{cc}
	\includegraphics[width=2.5in]{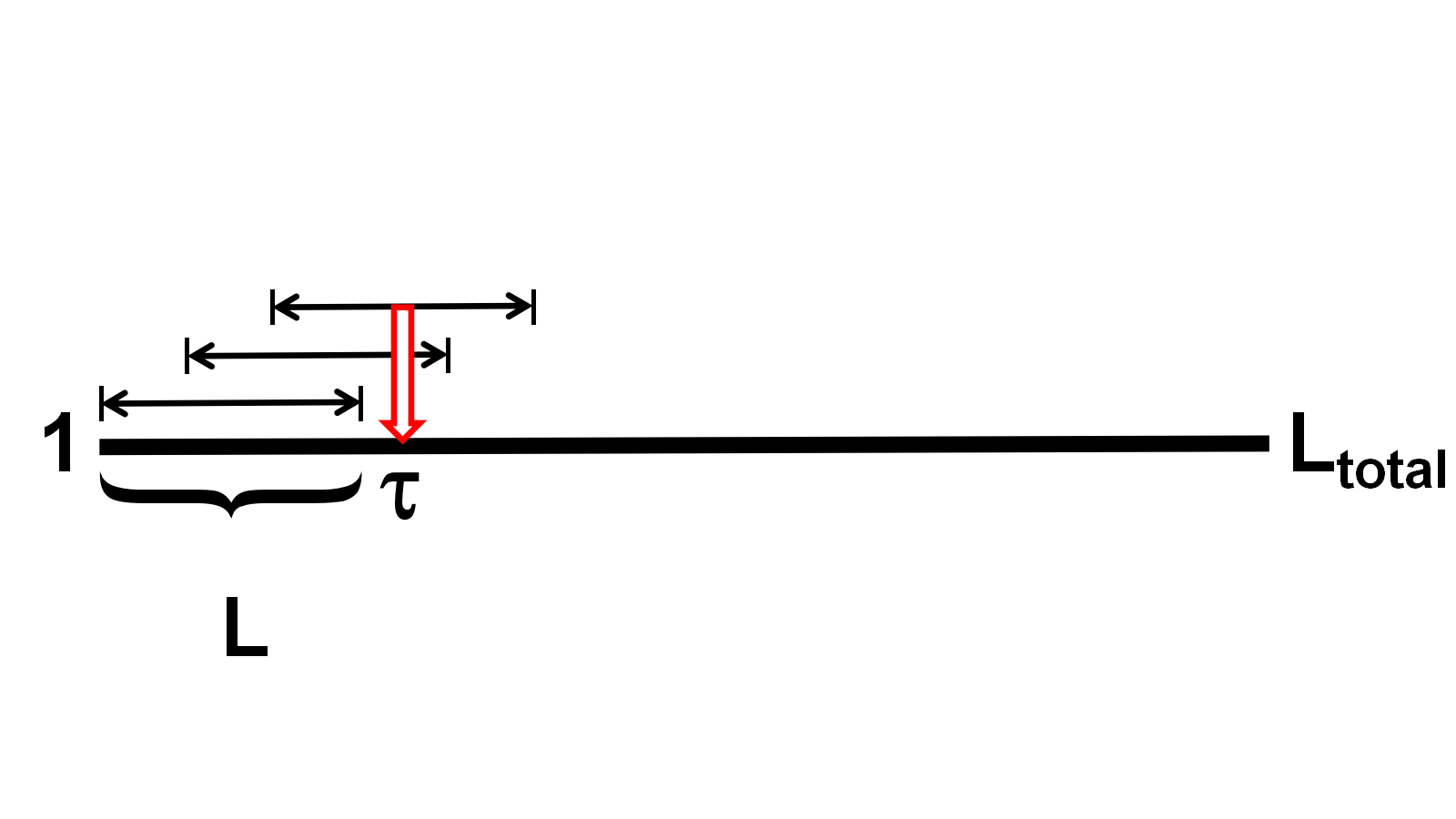} &
	\includegraphics[width=2.5in]{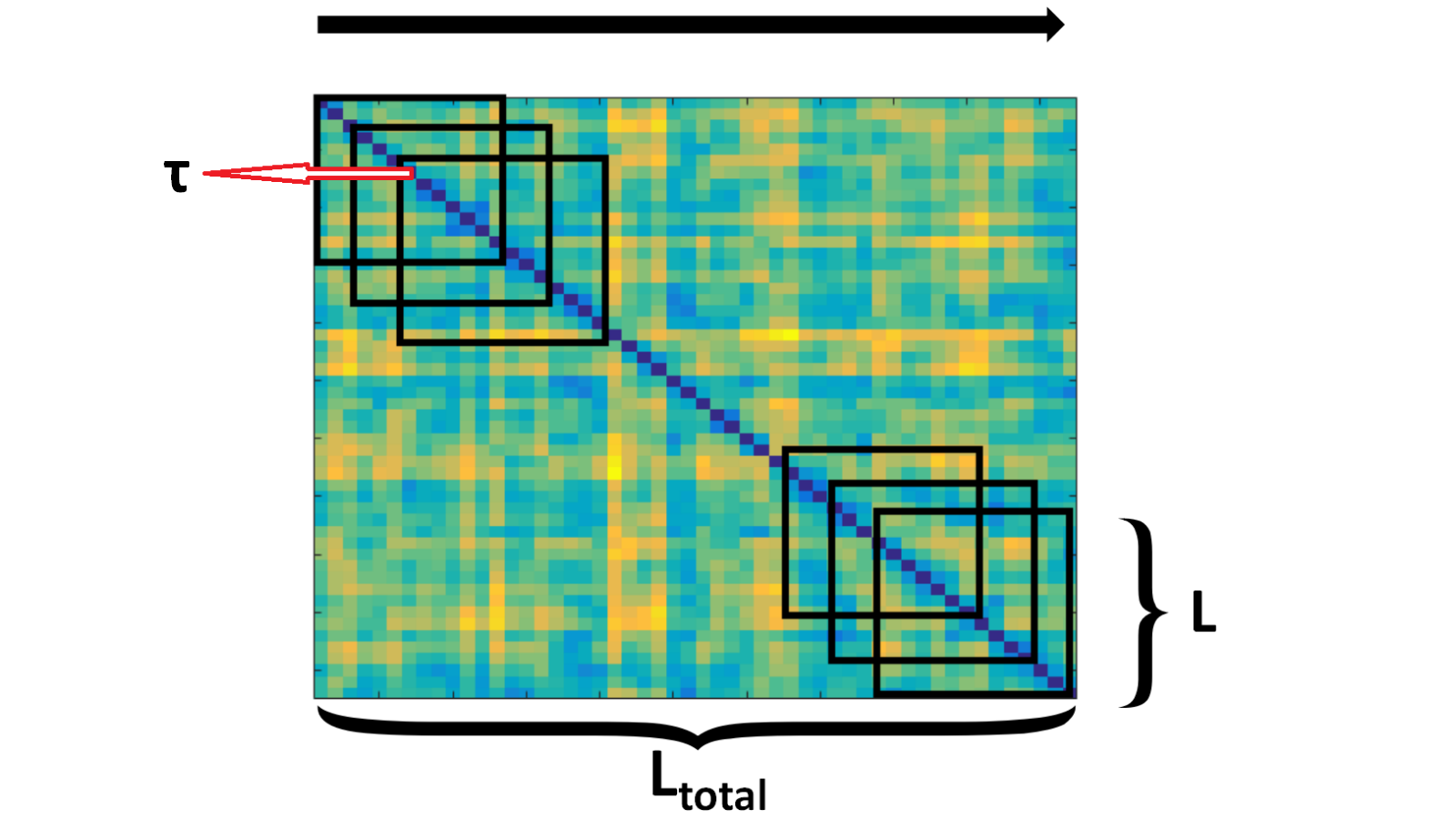} \\
	(a) & (b) \\ 
	\end{tabular} 
	\caption{(a) An example of a moving window embedded in the sequence of covariance matrices. (b) An example of a moving window embedded in the full distance matrix between all covariance matrices from time $1$ to time $L_{total}$. In each moving window, distance matrix  of length $L$ is used to form MST, and change-point test is performed at midpoint $\tau$ in the window. } \label{fig:distwindow}
\end{figure}

	Before we proceed further, we make the following observations about this MST-based change-point detection: 
	\begin{itemize}
		\item This method is theoretically derived for data points that are sampled independently from 
		their respective distributions. However, elements of the SPDM time series under study are seldom independent.  Later in the paper
		we present empirical studies that show that this method also works well for data with dependent structures. 
		
		\item When we are testing for  change-point at a time $\tau$, we assume that the samples before and 
		after $\tau$ are from the same distribution, respectively. In other words, there are no other change-points other than 
		$\tau$. 
		However, as we know, FC over large time intervals can potentially have multiple change-points. 
		We mitigate this issue by using a window-based approach: 
		we choose a large window of a fixed length, say $L$, and test 
		whether the midpoint of that window is a change-point or not. In this way, 
		we do not use the full time-series data for change-point detection at a time $\tau$, we only use data in the vicinity 
		of $\tau$. We repeat the process after moving the large window forward, as shown in Figure~\ref{fig:distwindow}.

		\item 
		This graph-based approach is not suitable for detecting change-points at the beginning and at the end of the time series. Therefore, 
		we simply avoid testing these boundary points in our experiments. This is reasonable because the concept of change points
		is valid only away from the boundaries. 
		
		%%% An example of moving window embedded in the distance matrix is given in Fig. \ref{fig:windowtodetectchange}.

	\end{itemize}

	%\iffalse
	%\begin{figure}
	%\begin{center}
	%\begin{tabular}{|c|}
	%\hline 
	%\includegraphics[width=2.6in]{figs/mst,19points,CP=10.jpg}  \\
	%\hline
	%\includegraphics[width=3.1in]{figs/mst,39points,noCP.jpg}  \\
	%\hline
	%\end{tabular}
	%\caption{Top: MST involving 19 SPDM matrices with a real change-point at  $\tau =10$. The two groups are
	%$\{1,2,\dots, 9\}$ and $\{10, 11,\dots,19\}$. 
	%Bottom: MST involving 39 SPDMs  with no change-point. The groups are formed by randomly 
	%dividing the points in two groups. The red edges represent connections 
	%across the chosen groups.}
	%\label{fig:graph-1}
	%\end{center}
	%\end{figure}
	%\fi

	%\begin{figure*}
	%\begin{minipage}[t]{0.5\linewidth}  
	%\includegraphics[width=6 cm]{figs/mst,19points,CP=10.jpg}
	%\label{fig_rescheduled-timetable_sample} 
	%\end{minipage}
	%\hspace{1ex} 
	%\begin{minipage}[t]{0.5\linewidth}  
	%\includegraphics[width=8 cm]{figs/mst,39points,noCP.jpg} 
	%\label{fig_rescheduled-timetable_sample} 
	%\end{minipage}
	%\caption{Left: MST involving 19 SPDM matrices with a real change-point at  $\tau =10$. The two groups are
	%$\{1,2,\dots, 9\}$ and $\{10, 11,\dots,19\}$. 
	%Right: MST involving 39 SPDMs  with no change-point. The groups are formed by randomly 
	%dividing the points in two groups. The red edges represent connections 
	%across the chosen groups.}
	%\end{figure*}

	\subsection{Estimation of Connectivity Graphs for Displays}
	
\begin{figure}
		\begin{center}
			\begin{tabular}{ccc}
				\includegraphics[width=1.0in]{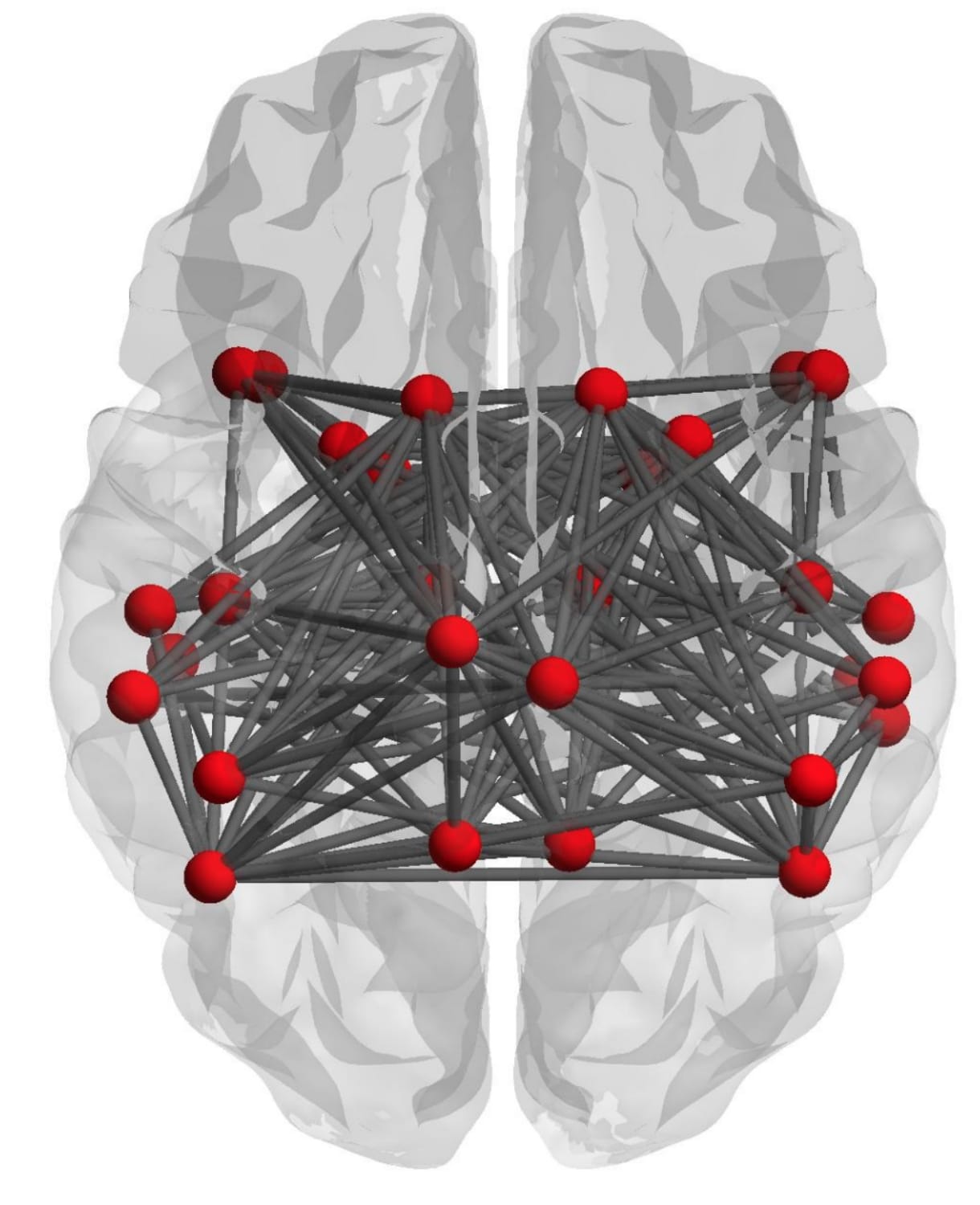}   &
				\includegraphics[width=1.0in]{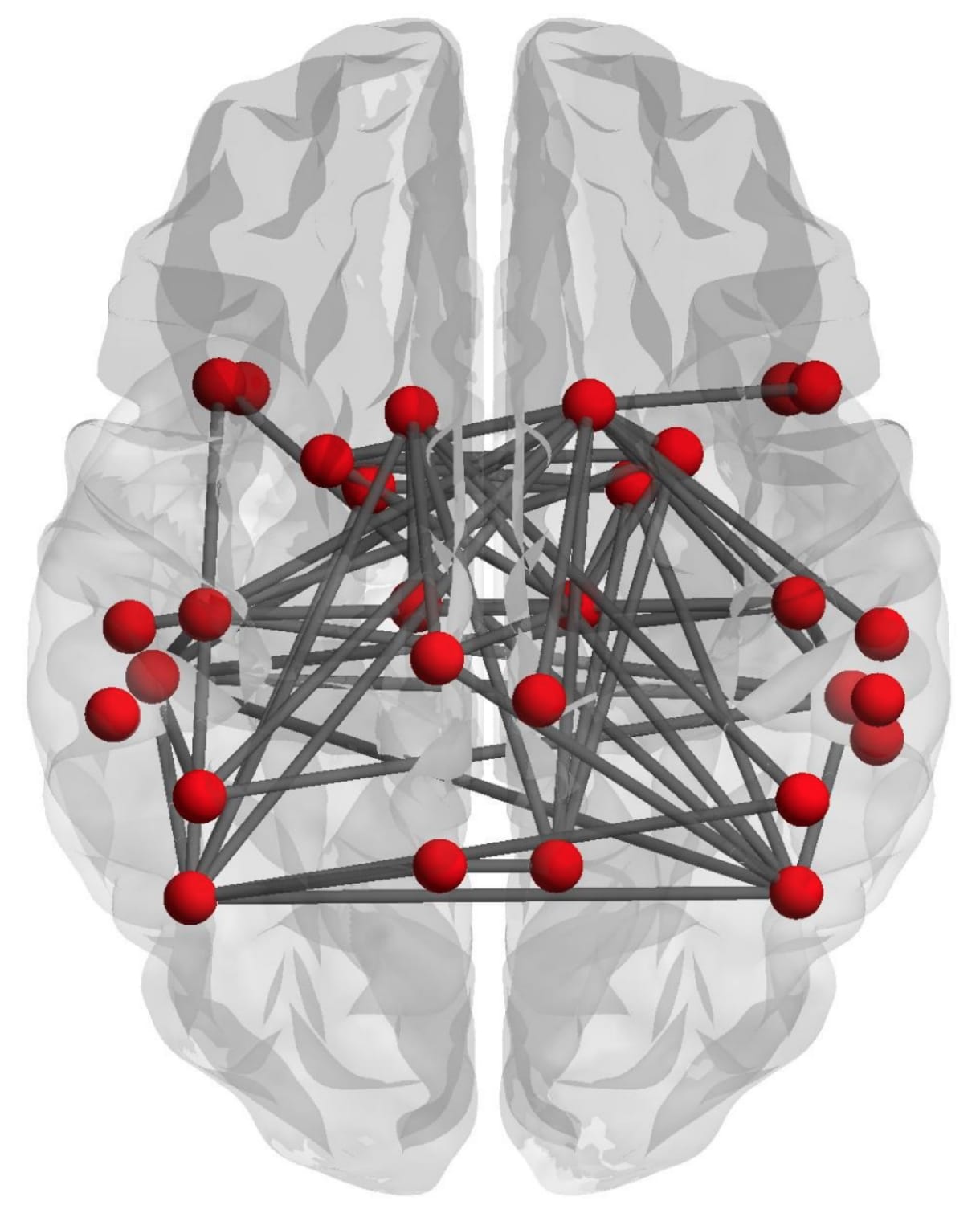}   &
				\includegraphics[width=1.0in]{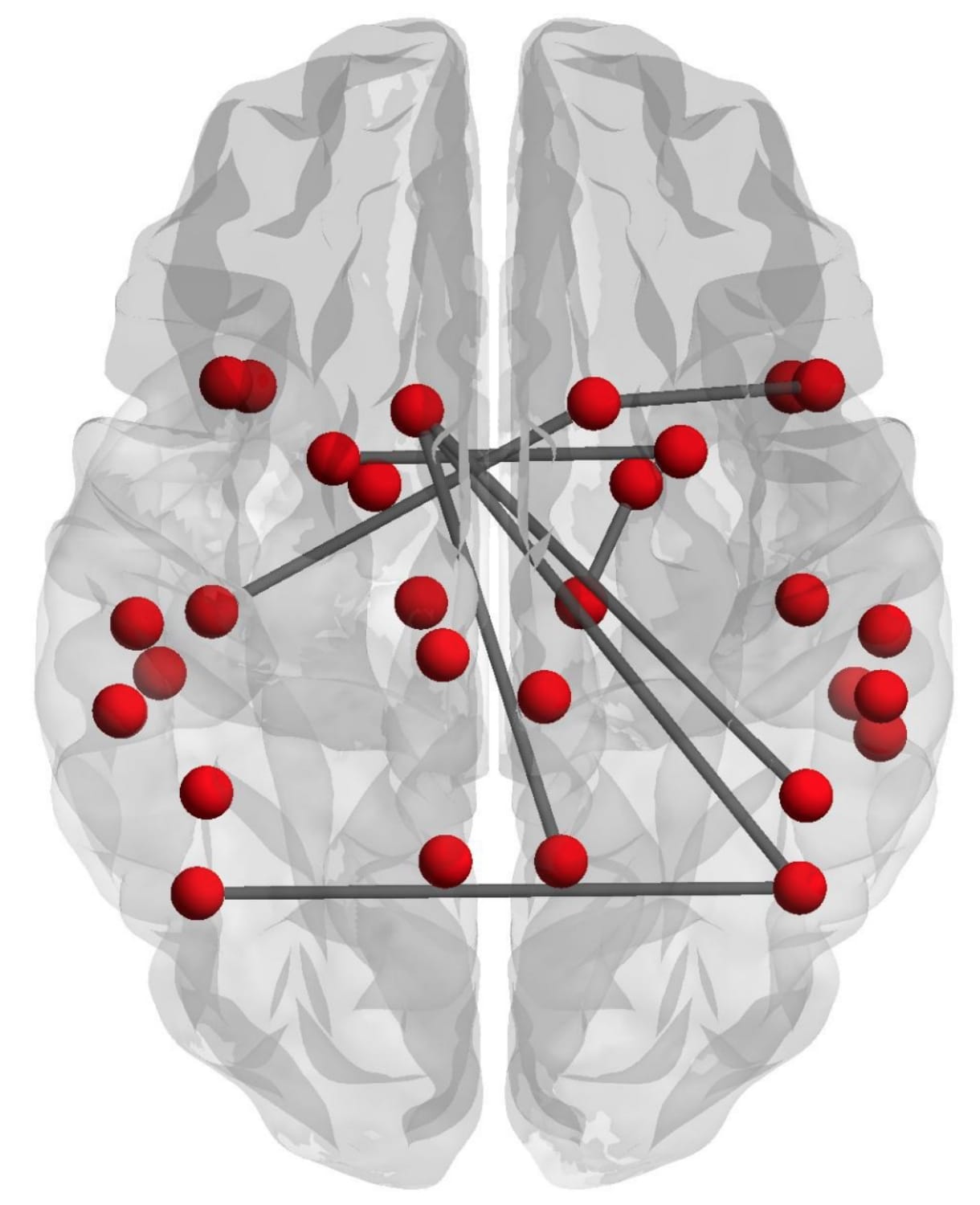}  \\
				$\rho$=0.01& $\rho$=0.03 & $\rho$=0.06 \\
			\end{tabular}
			\caption{ Estimated connectivity of 30 ROIs with different $\rho$s.} \label{fig:glasso}
		\end{center}
	\end{figure}

	For the purposes of display, one often wants to estimate a graph associated with a FC status. This helps visualize 
	connectivities across regions in a more precise fashion. One needs this, for example, to display 
	brain networks associated with FC over short intervals.  Since we are representing FC via a covariance matrix, we
	face the problem of estimating a network or graph from a covariance matrix. 
	We accomplish this using graphical Lasso (\cite{Friedman01072008}), as follows.
	Let the full observation period be divided into sub-intervals separated by the detected change-points 
	$\tau_1$, $\tau_2$, etc. For each sub-interval $[\tau_j, \tau_{j+1})$ we compute the Karcher 
	mean of all SPDMs over this sub-interval, under the Riemannian structure described earlier; call this mean $\bar{P}_j \in {\cal P}$. 
	Then, we use this mean SPDM as a representative FC of that subinterval and estimate the associated
	precision matrix  using: 
	$$
	\hat{\Theta}_j = \underset{\Theta \ge 0}{\operatorname{argmin}}\left(\mbox{tr}(\bar{P}_j \Theta) - \log (det(\Theta)) + \rho \sum_{i \ne k}{|\Theta_{ik}|}
	\right),\ 
	$$
	where $\rho>0$ is a parameter controls density of the generated graph (Figure~\ref{fig:glasso} illustrates 
	the effect of $\rho$ on the resulting graph). Given the estimated precision matrix $\hat{\Theta}_i$, the adjacency matrix is obtained
	by a simple thresholding, i.e. 
	$(A_j)_{ik} = 1_{|\hat{\Theta}_{ik}| > \epsilon=0.001}$. This adjacency 
	matrix is then used to represent the undirected graph formed by the corresponding brain regions. We emphasize that this graph estimation 
	is simply for display purposes and has no bearing on the change-point patterns that we study later. 
	
	We summarize the overall procedure for detecting change-point patterns in FC observations. 
	\begin{algo}{\bf Build Stationary Connectivities from fMRI Data} \label{algo:overall_procudure}
		\begin{enumerate}
			
			\item Use {Algorithm} \ref{algo:change-point-detect}, to form a normalized test statistic $Z(\tau)$, for each $\tau \in \{0.5L+1, 0.5L+2, \dots, L_{total}-0.5L+1\}$.  Make a list of $\tau$s such that $Z(\tau) > 3$; call them $\{\tau_j\}$. 
			(If several successive $\tau$'s qualify, then we select only the local maximizers of $Z$; see Remark 2.)
			These selected $\tau$'s denote the change-points of FC over the interval $[0,T]$.
			
			\item  For each sub-interval $\{\tau_j, \tau_{j+1}\}$,  find the Karcher 
			mean of  covariance matrices $\{ P_{\tau_j}, \dots, P_{\tau_{j+1}}\}$
			to form $\bar{P}_j$.  Use the mean SPDM to estimate the precision matrix and the adjacency matrix, which ultimately forms the brain network for that interval.
			
		\end{enumerate}
	\end{algo}
	
	\section{Temporal Synchronization of Change-Point Statistics}
	
	Algorithm \ref{algo:overall_procudure} enables us to detect change-points and estimate connectivity graphs
	associated with homogeneous sub-intervals, for each subject individually.  
	Additionally, it provides us with the test statistic $Z(\tau)$ as a probabilistic measure of 
	having a change-point at $\tau$, for each $\tau$ of interest over the interval $[0.5L+1,L_{total}-0.5L]$. 
	Next we are interested in studying patterns of change-points associated with the dynamic FC of a 
	group of individuals when they perform identical or similar tasks. The question is:  Are there any change-point patterns that are common across subjects, 
	under such similar experimental conditions? It turns out that simply visualizing $R(\tau)$ for a number of individuals does not lend
	to any apparent patterns. The reason is that 
	different individuals can have different reaction times for the same external stimulus. 
	Also, in any practical scenario, it is often hard to perfectly align all stimuli. The algorithm 
	used in change-point detection also introduces some variability in detected points. Finally, the randomness of data also leads to variability in 
	strengths and locations of the detected peaks in the test statistic. Together, these factors results in large variability in change point 
	detection results across subjects, reducing the possibility of finding patterns by naked eye. 
	However, if we take into account the temporal variation across individuals, 
	and align these functions to reduce inter-subject variability, a pattern emerges. 
	
	To enable temporal registration of test statistics across subjects,  we first fit a smooth, 
	densely-sampled function to the discrete $Z(\tau)$ observations of each subject, using cubic splines.
	Then, we use an elastic method, based on the classical Fisher-Rao metric, to obtain the warping functions that 
	best align these fitted splines across subjects (\cite{srivastava-klassen:2016}). 
	Let $\{f_i, i=1,\dots,m\}$, for $m$ subjects,  be the set of fitted smooth functions on the observed time interval, e.g. $[0,L_{total}]$.  
	Our goal is to find a set of time warping functions $\{\gamma_i\}$, such that $\{f_i \circ \gamma_i\}$ are 
	temporally aligned as well as possible. Let $\Gamma$ be the set of all positive
	diffeomorphisms on $[0,L_{total}]$ such that $\gamma(0) = 0$ and $\gamma(L_{total}) = L_{total}$; its elements play the role of time warping functions.
	Let $q$ be the {\it square-root slope function} (SRSF) of the function $f$, defined as  
	$q(t)= \text{sign}\{\dot{f}(t)\} \sqrt{{|\dot{f}(t)|}}$. (For a detailed description of SRSFs and their use in functional alignment please 
	refer to Chapt 4. of \cite{srivastava-klassen:2016}.) To jointly register functions $\{f_1,...,f_m\}$, the 
	iterative procedure is as follows: initialize a template $\mu_q$ and iteratively solve for 
	\begin{equation}
	\gamma_i = \arg \inf_{\gamma \in \Gamma} \| \mu_q - (q_i \circ \gamma) \sqrt{\dot{\gamma}} \|, i=1,2,\cdots,m, \text{and} \ \ 
	\mu_q = {1 \over m} \sum_{i=1}^m (q_i \circ \gamma_i) \sqrt{\dot{\gamma}_i} \quad. 
	\end{equation}
	Each optimization over $\Gamma$ can be solved using the dynamic programming algorithm. We present the alignment algorithm below: 
		
		\begin{algo}\label{algo:alignment} {\bf Temporal Alignment of Change-Points}
		
			\begin{enumerate}   
				\item Given a set of functions $f_1, f_2,... f_m$ (smooth functions fitted to the test statistics $Z(\tau)$) 
				on $[0,L_{total}]$, let $q_1,q_2,...,q_m$ denote their SRSFs, respectively. 
				
				\item Initialization step: Select $\mu_q = q_j$, where $j$ is any index in $\argmin_{1 \leq i \leq m} \Arrowvert q_i - {1 \over m} \sum_{k=1}^m q_k\Arrowvert$ .
				
				\item For each $q_i$ find $\gamma_i$ by solving: $\gamma_i = \arginf_{\gamma\in\Gamma} \Arrowvert \mu_q- (q_i \circ \gamma) \sqrt{\dot{\gamma}})\Arrowvert$. This alignment step is performed using dynamic programming algorithm.
				 
				\item Compute the aligned SRSFs $\widetilde{q_i} = (q_i \circ \gamma_i) \sqrt{\dot{\gamma}}$.  
				
				\item If the increment $\Arrowvert {1 \over m} \sum_{i=1}^m \widetilde{q_i} - \mu_q \Arrowvert$ is small, then stop. Else, update the mean using $\mu_q = {1 \over m} \sum_{i=1}^m \widetilde{q_i}$ and return to step $3$.
				
				\item Compute the mean $\overline {\gamma} = {1 \over m} \sum_{i=1}^m \gamma_i $. Center the estimate $\mu_q$ with respect to the set $\{q_i\}$, 
				according to $\widetilde{\mu_{q}} = (\mu_q \circ \overline {\gamma}^{-1}) \sqrt{\dot{\overline {\gamma}}^{-1}})$.  
				
				\item For i = 1,2,...,m, find $\gamma_i$ by solving $\gamma_i = \arginf_{\gamma\in\Gamma} \Arrowvert \widetilde{\mu_{q}}- (q_i \circ \gamma) \sqrt{\dot{\gamma}})\Arrowvert$. Compute the aligned functions as $\widetilde{f_i} = (f_i \circ \gamma)$.

			\end{enumerate}
		\end{algo}

	%After the alignment, we have the following set of functions representing change-point characteristics. 
	To demonstrate the effects of alignment, we start with the original test statistic functions $\{Z_i\}$ and threshold them to obtain the 
	change-points $\{C_i\}$; $\{C_i\}$ is a set of binary functions on the interval $[0,L_{total}]$,
	$1$ at the change-points and $0$ everywhere else. The smooth functions fitted to $\{Z_i\}$ are termed $\{f_i\}$. From Algorithm \ref{algo:alignment}, we obtain the optimal warping functions
	 $\gamma_i's$ to align $f_i's$. The aligned functions are denoted by $\{\tilde{f}_i = f_i \circ \gamma_i\}$ and the aligned change-points are given by 
	 $\{ \widetilde{C}_i = C_i \circ \gamma_i\}$. The $\widetilde{C}_i $'s  are then used to visualize and discover the change-point patterns across subjects.

	\section{Experimental Results}
	In this section we demonstrate our framework using both simulated and real fMRI datasets taken from the Human Connectome Project (HCP) database  (\cite{hcp2013}). 
	
	\subsection{Simulation Study}
An important use of simulated data analysis is to evaluate 
methodology in situations where the ground truth is known and can be 
compared against the quantities estimated from the data. In this section, we simulate data from some simple models that mimic real data settings.
We use the popular VAR (Vector Auto Regression) model, in two different state spaces and with a variety of parameter values, in order to 
study several different scenarios. 
Since our framework detects change points in covariance trajectories, we can simulate the data either in: (1) the original fMRI signal space and then 
form covariance trajectories from that simulated data, or (2) the covariance space directly. We will study both the cases in this section. 
Additionally, we will choose a variety of model parameters in both cases to test the algorithms from different perspectives. 
Specifically, we will use parameters estimated from the real HCP data and also some randomly selected values to make the evaluation 
quite comprehensive. With these data settings we demonstrate the success of our method under different parameter choices (window size $W$, step 
size $S$, larger window $L$, etc) involved in the 
algorithms. 

In terms of the models, we make the following two broad choices: 
\begin{enumerate}
\item {\bf VAR Model in Signal Space}: Here we model the data in the original fMRI signal space, generating 
a time series $\{X_t \in \real^n, t =1 , 2, \dots, 300\}$, where $n$ is the number of ROIs
used  in the experiment. The VAR(1) model for 
this case is: $X_{t+1} = A_t X_t + B_t {\cal E}_t$, where $A_t, B_t \in \real^{n \times n}$ are time-varying model parameters and 
${\cal E}_t \in \real^n$ is the noise. {\bf In all simulation experiments except 
Experiment 1, we scale the data at each time $t$ by dividing the sum of values in $X_t$ 
at all dimensions, i.e., ${X_t}^{(k)} = {X_t}^{(k)} / (\sum_{j=1}^n {X_t}^{(j)})$. One can use other 
methods to normalize such simulated time series data also (\cite{PMID:27428217}).   
An example of simulated data in Experiment 1 is shown in Figure~\ref{fig:data_scale}. 
One can see that the general scale of values appears similar to the real fMRI data (an example is shown in Figure~1)}.
\begin{figure} 
	\centering  
	\begin{tabular}{c}
		\includegraphics[width=3.5in]{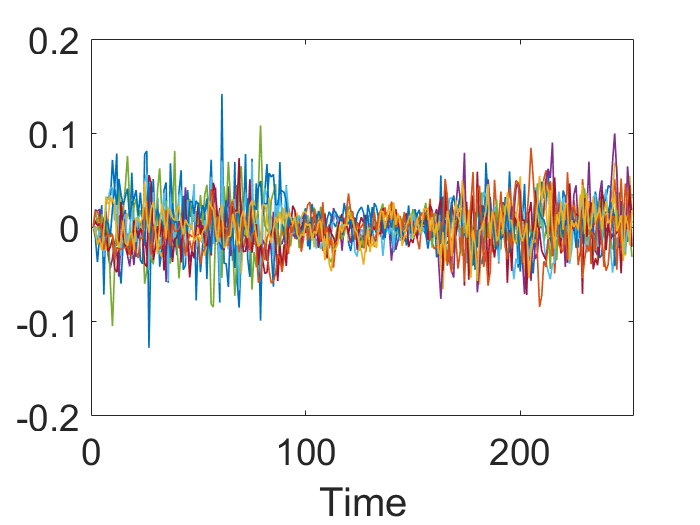} 
	\end{tabular} 
	\caption{{\bf An example of simulated data from VAR model in signal space when n = 10.}} \label{fig:data_scale}
\end{figure}
By choosing different values of $A_t$ and $B_t$, we explore a variety of scenarios. 
We deliberately introduce change-points in a time series by dividing the full observation internal 
into subintervals and by changing model parameters for different intervals.  
Given a simulated time series under this model, we then compute the corresponding covariance trajectory (using 
different values of algorithm parameters $W$, 
$S$,  $L$, and $L_{total}$) and study the problem of change-point detection using our approach. 
 In all simulated examples, except when estimating parameters from the real data,  we divide the whole time 
interval $[0,300]$ into three subintervals with change-points located at $100$ and $200$. 
Also, we use ${\cal E}_t \sim {\cal N}(0, 0.01*I)$ in the following experiments. 
{\bf Experiments 1-4} are based on data simulated from this data model. 

\item {\bf VAR Model in Covariance Space}: To add variety to the experimental studies, we
also model the time series in the covariance space directly, using 
a model of the type:  $P_{t+1} = (A_t + B_t {\cal E}_t) P_t (A_t + B_t {\cal E}_t)^T  \in \P$. Once again, $A_t, B_t \in \real^{n \times n}$
are model parameters and ${\cal E}_t \in \real^{n \times n}$ is a matrix of random observations. Different choices of $A_t$ 
and $B_t$ lead to different actual models for the data. We introduce change points in the data by 
dividing the total time  $[0,48]$ into subintervals, with change-points at $17$ and $33$, 
and use different parameters (but the same model) in different subintervals. 
Note that one can map this time scale to the time scale of the fMRI signal  using  $\tau \mapsto (\tau-1)S+W$, 
where $W=16$ and $S=6$ in most of our experiments. {\bf Experiment 5} follows this data model.  	

\end{enumerate}

In some cases, 
where we use random SPDMs as parameters in simulation models, first we randomly generate an $n 
\times n$ square matrix $A_0$, then we form an SPDM using 
$A_{0}{A_{0}}^T+nI_n$. In both models we make $A_{ij}$s and $B_{ij}s$ unit determinant to prevent  time series from going to infinity.  
Using these two model spaces (and five associated experiments) we perform a number of simulation studies. We simulate
the data numerous times, estimate the change-points and collate the results, as described next.

	\subsubsection{{\bf Experiment 1}: Simulation using Real Parameters}
\begin{figure*} 
		\centering
		\begin{tabular}{ccc}
			\includegraphics[width=1.6in]{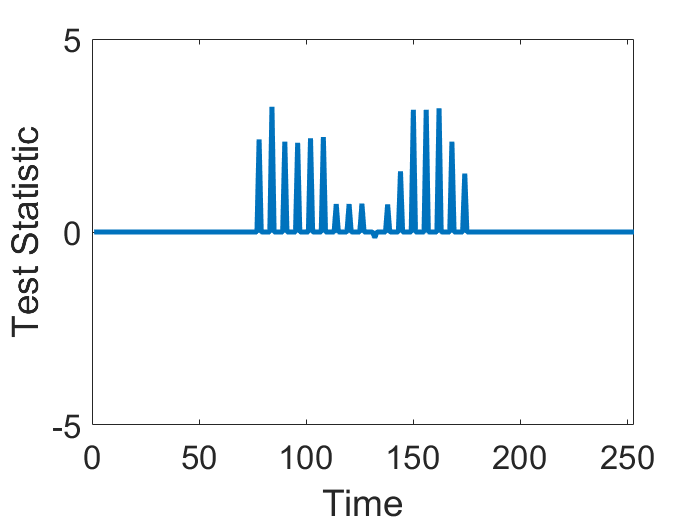}&
			\includegraphics[width=1.6in]{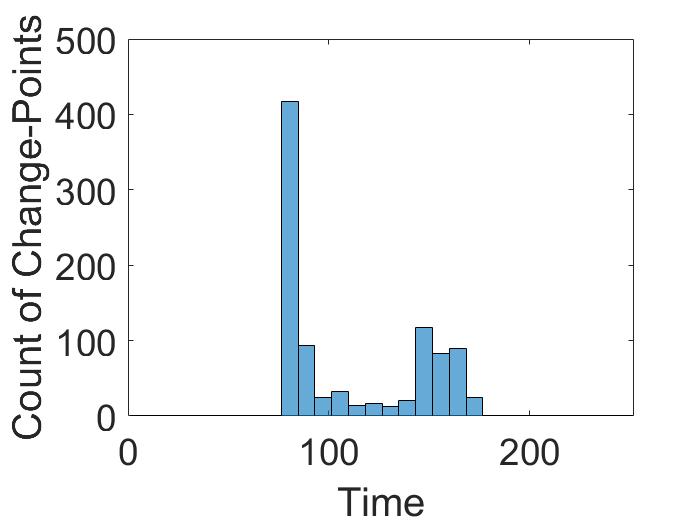}&
			\includegraphics[width=1.6in]{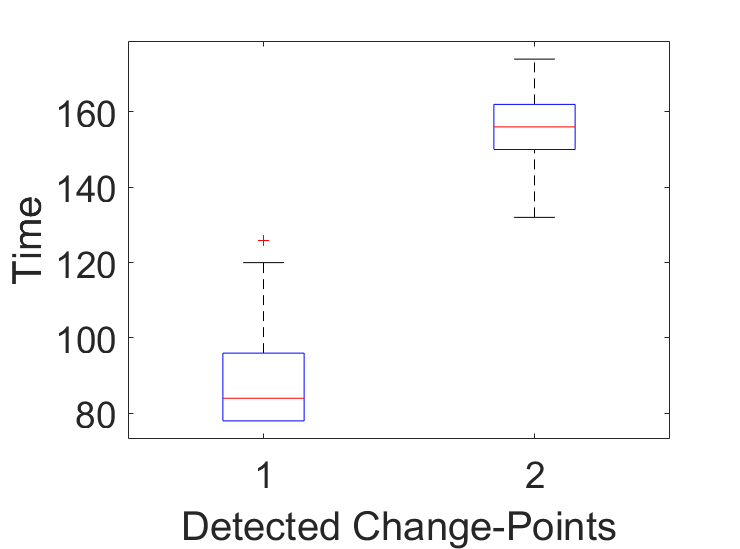} \\
			(a) & (b) &(c) \\
			\includegraphics[width=1.6in] {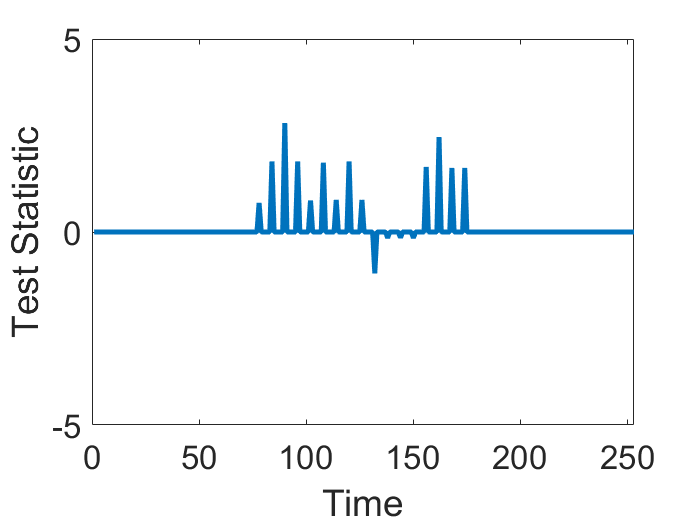}&
			\includegraphics[width=1.6in]{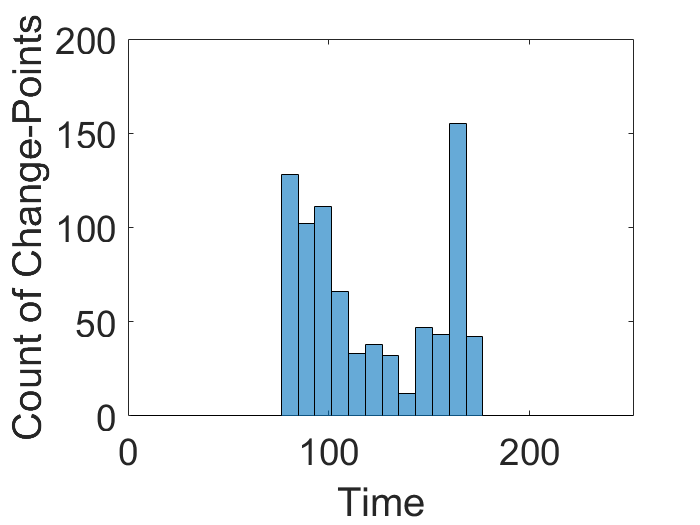}&
			\includegraphics[width=1.6in]{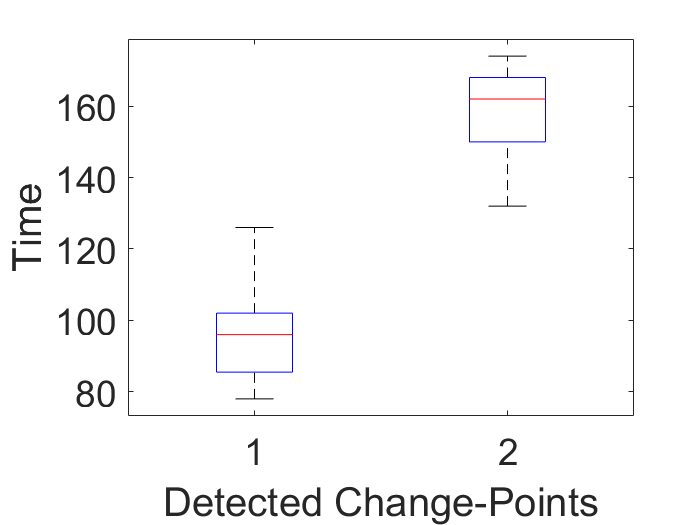} \\
			(d) & (e) &(f) \\						
		\end{tabular}
		\caption{ Results of {\bf Experiment 1}. Top: simulation from 10 ROIs in CinguloOperc network during gambling task. Change-points are located around time $t=84$
		and $156$.  Bottom: simulation from 10 ROIs in DorsalAttn network during gambling task. Change-points are located around time $t=90$ and $156$. 
		(a), (d) Temporal evolution of the test statistic on real data. (b), (e) Histogram of detected change-points in 1000 simulation runs. (c), (f) Boxplot of 
		these detected change-points. } \label{fig:sim0}
	\end{figure*}

	 In this experiment we estimate parameters from real data under the VAR(1) models, and 
		use them to simulate new data. That is, the time series over the $j^{th}$-subinterval, with $j=1,2,3$, is set to be
		$X_{t+1}=A_{j} X_t  +  \sigma {\cal E}_t$, where $A_{j}$s are estimated from HCP data and $\sigma = 10$. 
		In practice we tried a number of values of $\sigma$ and found the results to be stable with respect to this choice.
	The parameters $A_{j}$s are estimated from HCP task data as follows. 
	First we pick $n$ ROIs from single subject gambling task data to form a multivariate time series (in the signal space) and apply
	Algorithm~\ref{algo:change-point-detect} to detect change-points. We then estimate coefficients $A_{j}$s from a VAR(1) model in each sub-interval, 
	and use these estimated $A_{j}$s to simulate new data $\{X_t\}$ using the model mentioned above. 
	Given the simulated data, we implement 
	Algorithm~\ref{algo:change-point-detect} with parameters $W=16$, $S=6$, $L/2=12$, $L_{total}=40$ on this data and the results are noted. 
	We perform this process 1000 times and the estimation summaries (histograms and box plots) of the detected change points
	 are shown in Figure~\ref{fig:sim0}. 
	(Note that, for the histograms, we define the change-points to be only the local peaks of the test statistic, as mentioned in Remark 2.)
	For the box plots,  we use the full test statistics (above the chosen threshold of 3), to create a larger data for this summary, 
	
	The two rows in Figure~\ref{fig:sim0}  correspond to results from two different sets of ROIs (for the same subject and the same task). The top example uses 
	10 ROIs in CinguloOperc network during the performance of the gambling task. Note that the change-points estimated from the real data, or the ground truth,
	are located
	at  $t=84,156$. The bottom 
	example comes from 10 ROIs in DorsalAttn network and the ground truth change-points here are located at  
	$t=90,156$. 
	Figure~\ref{fig:sim0}(a) shows the evolution of the test statistic $Z(\tau)$ versus $\tau$ on the original real data, while panel (b) shows a histogram of detected test statistic
	peaks over 1000 simulation runs. The panel (c) shows the corresponding box plot. This last part requires manually separating the two sets of detected points in each run, to designate a change point as either
	first or second, and we have used $t = 130$ for 
	this separation. 
	Together, these two panels show that the 
	estimates are concentrated around true change-point locations. In panel (c) the medians of two change-points are around $t=84$ and 
	$t=156$, coinciding with the true underlying change-points. 
	Similarly, panels (d), (e), and (f) show results associated with the second example. In 
	panel (f) we see the medians of two change-points are located around $t=96,162$, quite close to true change-point locations.

	\subsubsection{{\bf Experiment 2}: Studying Effects of Window Size $W$} 
	 In this experiment the goal is to study the effect of the window size $W$ on 
		change-point detection. We simulate the time series over the $j^{th}$-subinterval, with $j=1,2,3$, according to
		$X_{t+1}=  A_{2j} X_t + B_{2j} {\cal E}_t$, where $A_{21}$ is a randomly-generated, unit-determinant SPDM; 
		$A_{22}=I; A_{23}=2A_{21}-nI; B_{21}=A_{21}+nI; B_{22}=A_{21}+nI; B_{23}=A_{21}+nI$. Once we have the simulated data, 
		we try different values of $W$ (while keeping other algorithmic parameters same) and study the results. 
		In each run, we create a time series with $300$ time points, with $100$ points in each subinterval in a $10$-dimensional space ($n = 10$). 
	Here we fix $S=6$, $L/2 = 12$, and try different values of $W$.
	 
	Figure~\ref{fig:sim1} shows a subplots of the test statistics $Z(\tau)$ versus $\tau$ for 12 different values of $W$. 
	 Note that due to the correlated nature of the time series data, the high-value of test statistics is not in isolation but is accompanied by a set of high values in its neighborhood.  
	 We, thus, find a peak of $Z(\tau)$ value by looking for a local maximum (see Remark 2), 
	 i.e. the test statistic is increasing before that value and decreasing after it. 
	 The detected change points for these cases are as follows, from top-left to bottom-right: 
	 $W=4, t=124; W=8, t=128, 206; W=12, t=204; W=16, t=112,208; W=20, t=118,206; W=24, t=120,204; W=28, t=124,208; W=40, t=130; W=44, t=116; W=48, t=126,198.$
	 The standard deviations of the two detected change-points, across these different $W$s, are 
	 found to be $5.92$ and $3.44$, respectively, which highlights the stability of proposed detection method within the chosen range of $W$s. 
	 We caution the reader that  $W$ should not be too large to smooth over subtle changes 
	 that are critical to assessments of temporal evolution of FC (as mentioned in Remark 1).
	
	\begin{figure}
		\centering  
		\includegraphics[width=6in]{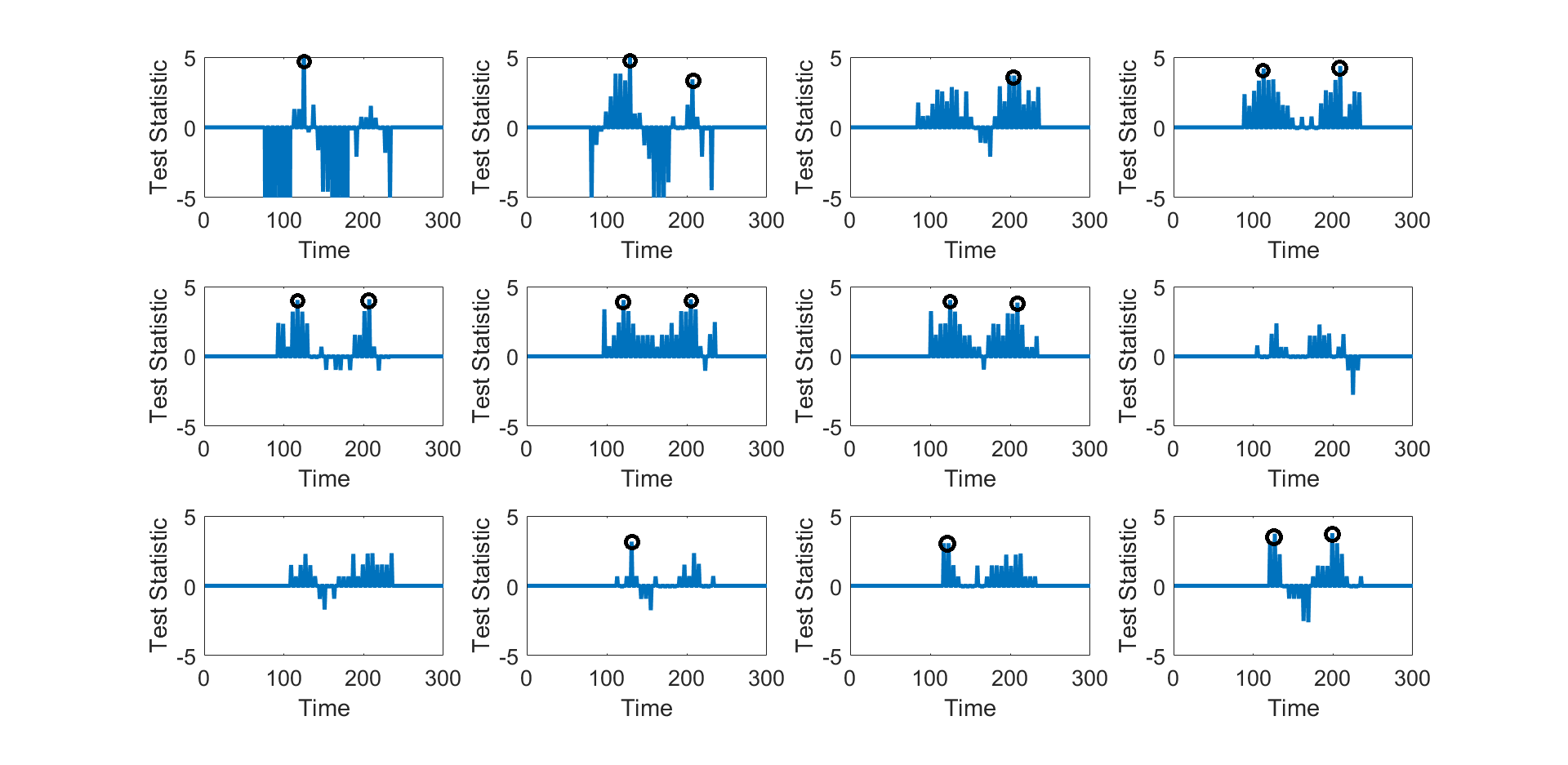}			
		\caption{ Results of {\bf Experiment 2}. Change-point detection on simulated data from Model 1.1, with window sizes $W=4, 8, 12, 16, 20, 24, 28, 32, 36, 40 , 44, 48$ from top left to bottom right. The standard deviations of change-point locations for the two change-points are $5.92$ and $3.44$.} \label{fig:sim1}
	\end{figure}
	
	\subsubsection{{\bf Experiment 3}: Studying Effects of Increasing $n$}
\begin{figure*} 
		\centering
		\begin{tabular}{ccc}
			\includegraphics[width=1.4in]{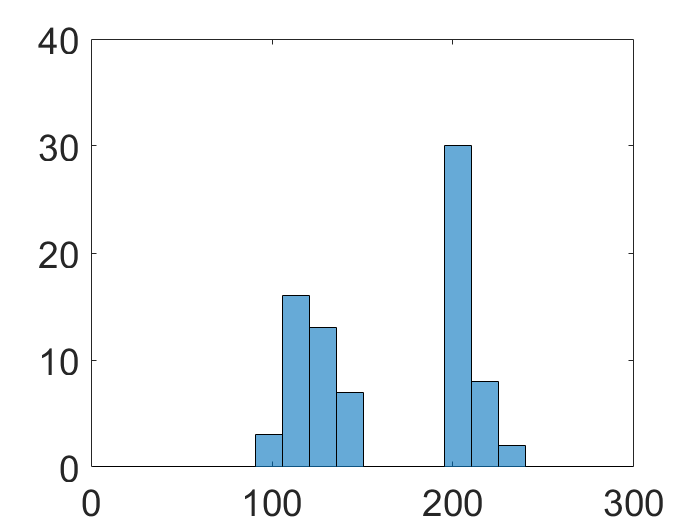}&
			\includegraphics[width=1.4in]{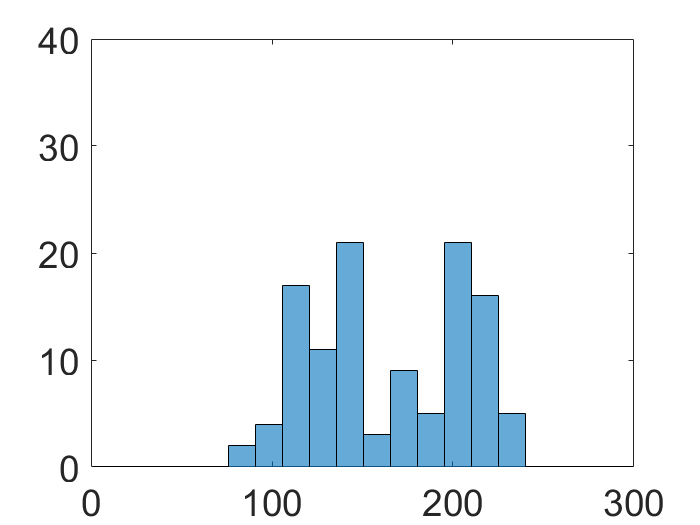}&
			\includegraphics[width=1.4in]{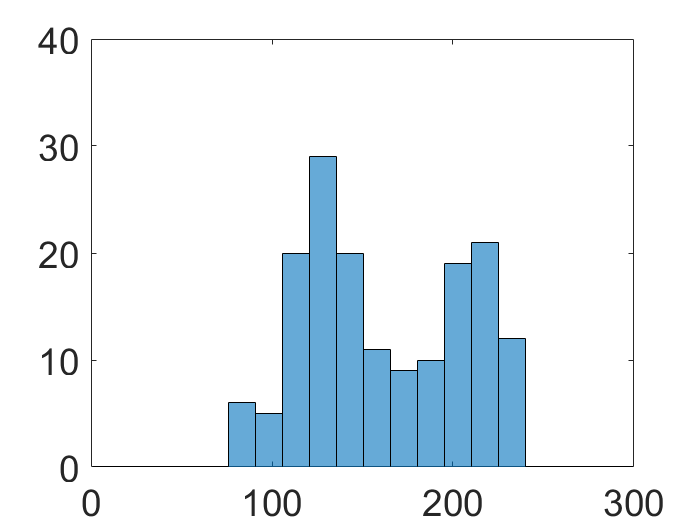} \\
			(Model 1.1, $n=4$) & (Model 1.1, $n=15$) &(Model 1.1, $n=50$) \\
%			\includegraphics[width=1.4in] {figs/simulation_VAR,model2,d=4,hist.png}&
%			\includegraphics[width=1.4in]{figs/simulation_VAR,model2,d=15,hist.png}&
%			\includegraphics[width=1.4in]{figs/simulation_VAR,model2,d=50,hist.png} \\
%			(Model 2, $n=4$) & (Model 2, $n=15$) &(Model 2, $n=50$) \\
			\includegraphics[width=1.4in] {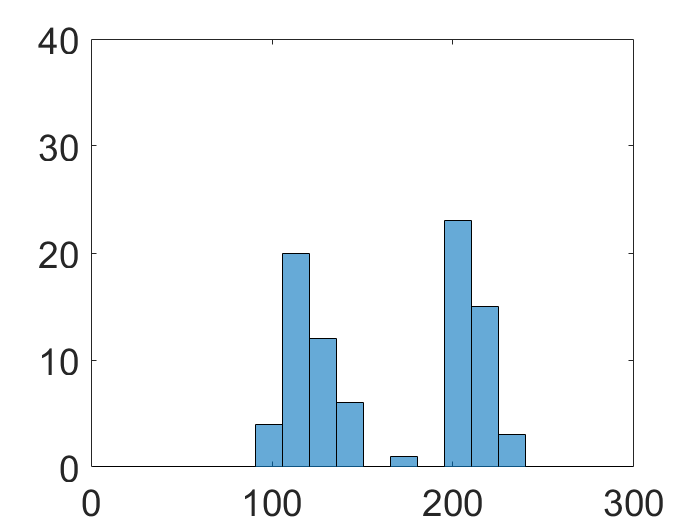}&
			\includegraphics[width=1.4in]{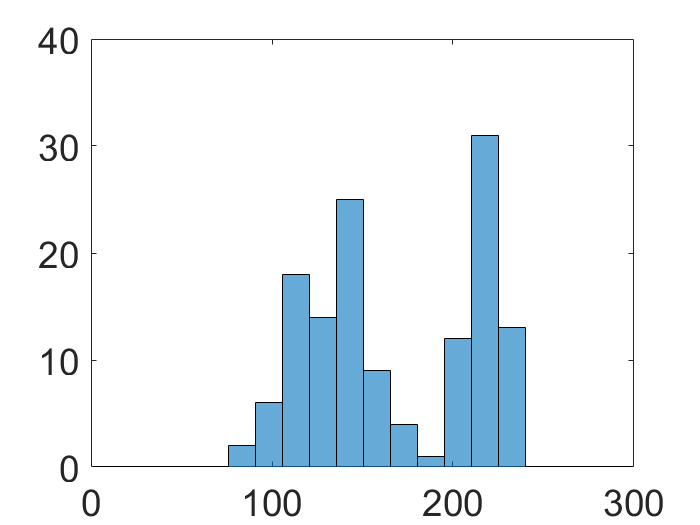}&
			\includegraphics[width=1.4in]{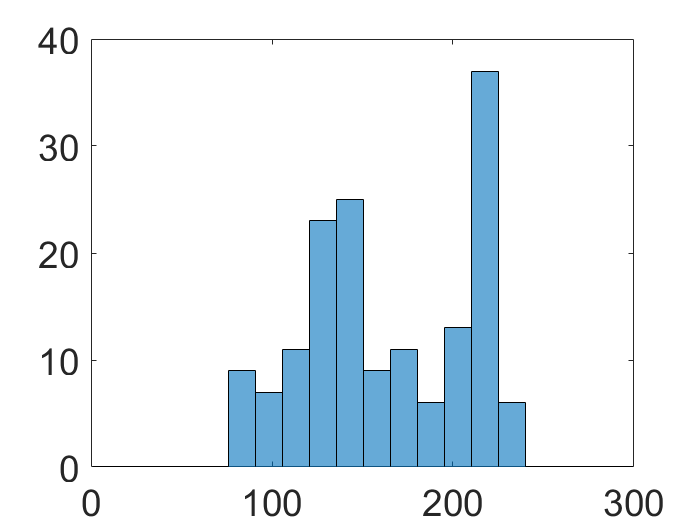} \\
			(Model 1.2, $n=4$) & (Model 1.2, $n=15$) &(Model 1.2, $n=50$) \\
			
		\end{tabular}
		\caption{ Results of {\bf Experiment 3}. Histograms of detected change-points given different models and dimensions with 100 runs in each case. Sample means and standard deviations for two change-points in Model 1.1 when n = 4, 15, 50 are (123.15, 11.69; 207.50, 8.88), (125.98, 17.16, 202.08, 20.45), (124.25, 16.35; 200.07, 24.23) respectively; Sample means and standard deviations for two change-points in Model 1.2 when n = 4, 15, 50 are (120.43, 12.70; 210.43.50, 11.61), (126.66, 16.48, 207.54, 24.44), (123.16, 19.40; 202.93, 24.14) respectively. } \label{fig:sim2}
	\end{figure*}

	The goal in this experiment is to study the effects of increasing the dimension $n$ on detection results, while 
		keeping the other parameters fixed. To accomplish that, we
		 simulate time series from two sets of model parameters: 
		 \begin{itemize}
		 \item One is the same as in $\bf Experiment$ $\bf 2$, call it Model 1.1. 
		 \item The other is 
		$X_{t+1}= A_{3j} X_t  + B_{3j} {\cal E}_t$, call it Model 1.2, 
		where $A_{31}$ is a randomly generated, unit determinant SPDM; $A_{32}=I; A_{33}=2A_{31}+nI; B_{31}=nI; B_{32}=I; B_{33}=J-I$.
		Here $J$ is an $n \times n$ matrix of all ones. 
		\end{itemize}
	For each choice of model parameters and $n$, we run the program 100 times and 
	collect the results. Figure~\ref{fig:sim2} shows histograms of detected change-points in each of the settings mentioned in the captions. 
	
	We observe that when $n$ is either small or moderate,  the algorithm can successfully detect change points with a high
	degree of accuracy. 
	However, when $n$ gets large, as 
	for example in the case of Model 1.1, $n=50$, the detection algorithm does not perform as well. This is because one needs a larger 
	sample size for estimating larger covariance matrices at the same precision level. 
	This issue can perhaps be mitigated by using different window parameters, or by introducing a dimension reduction technique, but it has not 
	been explored in this paper. 
	
	\subsubsection{{\bf Experiment 4}: Studying Effects of Increasing Temporal Correlations}
	In this experiment, we study the effects of an increase in temporal correlations of the time 
		series on the detection performance. We use the same model parameters as in ${\bf Experiment\ 2}$, 
		except that we replace $X_t$ with $(\sum_{i=1}^p X_{t-i+1})/k$, making it a VAR($p$) model. Increasing $k$ increases the 
		correlations in $X_t$s across time. 	
		
\begin{figure*} 
		\centering
		\begin{tabular}{ccc}
			\includegraphics[width=1.4in]{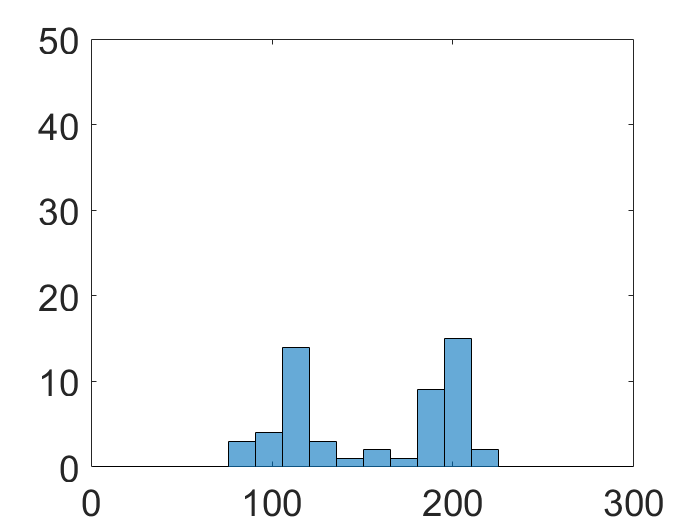}&
			\includegraphics[width=1.4in]{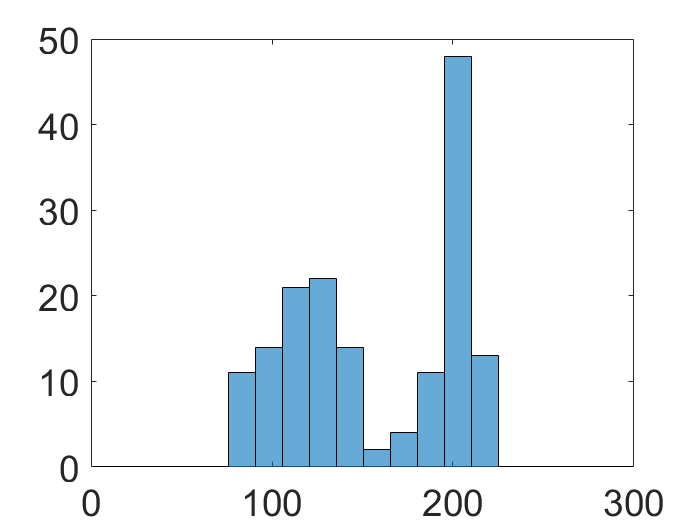}&
			\includegraphics[width=1.4in]{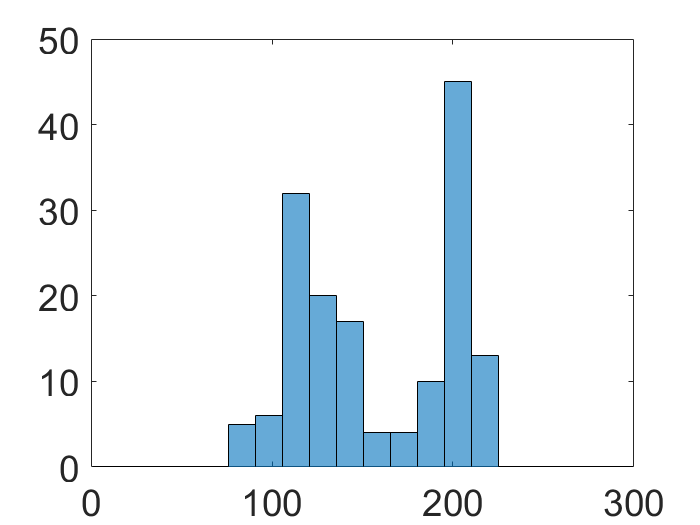} \\
			($p=3$, $n=4$) & ($p=3$, $n=15$) &($p=3$, $n=50$) \\
			\includegraphics[width=1.4in]{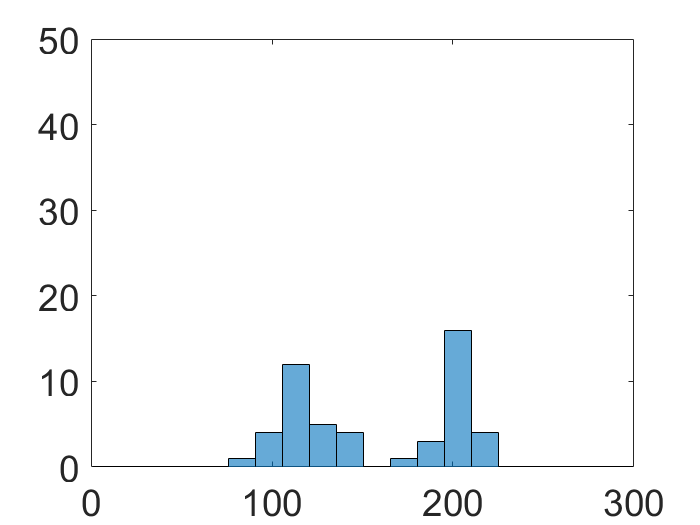}&
			\includegraphics[width=1.4in]{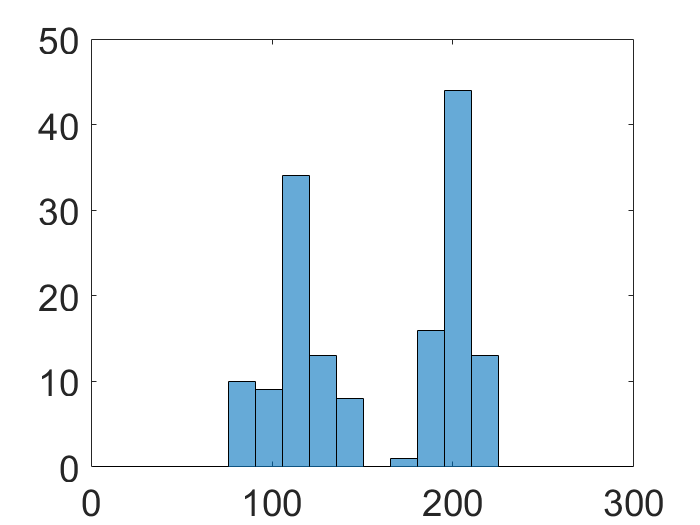}&
			\includegraphics[width=1.4in]{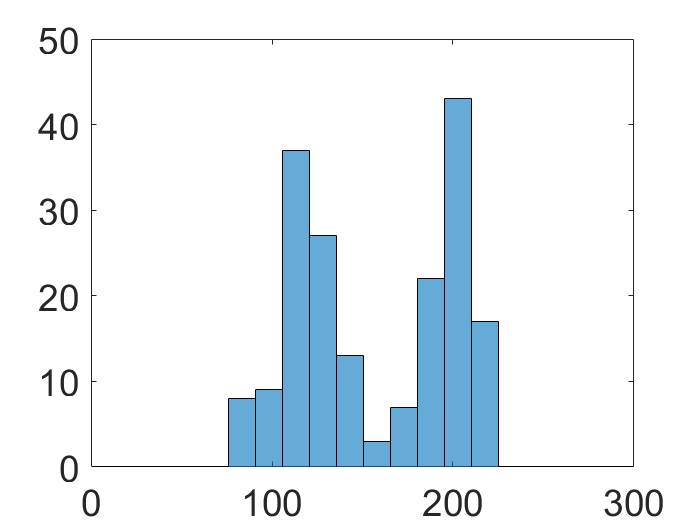} \\
			($p=5$, $n=4$) & ($p=5$, $n=15$) &($p=5$, $n=50$) \\
			
		\end{tabular}
		\caption{ Results of {\bf Experiment 4}. Histograms of detected change-points given VAR(3) and VAR(5) models with different dimensions with 100 runs in each case. Sample means and standard deviations for two change-points when $p = 3, n = 4, 15, 50$ are $(112.56, 14.52)$
		$(196.14, 13.80)$, $(116.05, 19.58)$, $(201.85, 12.90)$, $(120.23, 16.85; 200.92, 14.88)$ respectively; Sample means and standard deviations for two change-points in Model 1.2 when 
		$p = 5, n = 4, 15, 50$ are $(117.23, 14.62; 203.25.50, 9.58)$, $(113.35, 17.05, 203.51, 10.96)$, $(118.60, 16.16; 199.77, 14.22)$ respectively. } \label{fig:sim3}
	\end{figure*}

	Although the null hypothesis of change-point detection requires independent samples, according to \cite{chen2015}, the method still seems 
	quite effective when the data points are weakly dependent. 
	$\bf Experiment$ $\bf 1-3$ use VAR($p$) models with $p=1$, and the state at time $t$ only depends on the state at time $t-1$. 
	Now we increase dependence using VAR(3) and VAR(5) models and study the effects of this 
	dependence on the results. In a VAR(3) model, $X_t$ depends on $X_{t-1}, X_{t-2}, X_{t-3}$, and in a VAR(5) model $X_t$ depends on 
	$X_{t-1}$ to $X_{t-5}$. Figure~\ref{fig:sim3} shows histograms of detected change-points in $p=3$ and $p=5$ cases with different values of $n$ and using 100 runs in each setting.
	 
	Despite different degrees of dependencies in the data from case to case, our change-point detection algorithm 
	is quite consistent in detecting the original change-points. 
	Even when the dimension gets high, i.e. $n=50$, the process shows a high level of success.
	
	\subsubsection{ {\bf Experiment 5}: Direct Simulations of Covariance Sequences}
	In this case, we simulate the time series in ${\cal P}$ directly from two sets of parameters as follows: 
	\begin{itemize}
	\item	 Model 2.1: Using $P_{t}=(A_{4j}+  {\cal E}_t)P_{t-1}({A_{4j}+ {\cal E}_t)}^T$, where $A_{41}$ is a randomly generated SPDM; 
	$A_{42}=I;A_{43}=J-I$; 			
	\item  Model 2.2: Using the same model but different parameters, we have: 
		$P_{t}=(A_{5j}+  {\cal E}_t) P_{t-1}({A_{5j}+  {\cal E}_t)}^T$, 
		where $A_{51}=I$, $A_{52}$ is a randomly generated SPDM, 
		and $A_{53}=J-I.$
		As earlier, $J$ is an $n \times n$ matrix of ones and  ${\cal E}_t \sim {\cal N}(0, 0.01*I)$.
	\end{itemize} 
		
\begin{figure*} 
		\centering
		\begin{tabular}{ccc}
%			\includegraphics[width=1.4in]{figs/simulation_VAR_traj,model1,d=4,hist.png}&
%			\includegraphics[width=1.4in]{figs/simulation_VAR_traj,model1,d=15,hist.png}&
%			\includegraphics[width=1.4in]{figs/simulation_VAR_traj,model1,d=50,hist.png} \\
%			(Model 6, $n=4$) & (Model 6, $n=15$) &(Model 6, $n=50$) \\
			\includegraphics[width=1.4in]{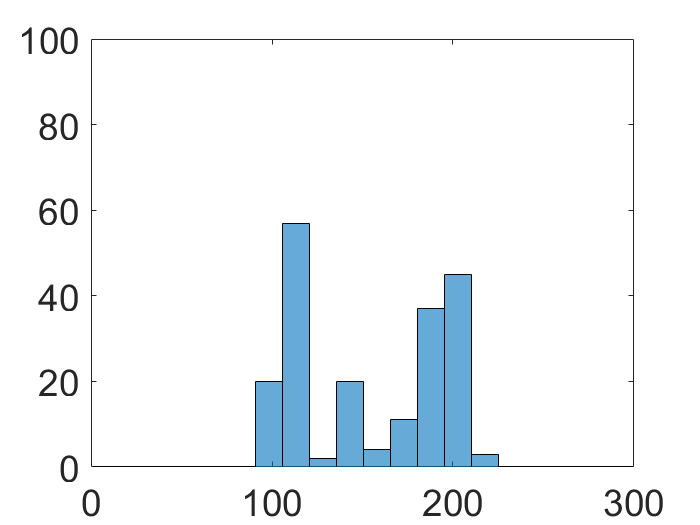}&
			\includegraphics[width=1.4in]{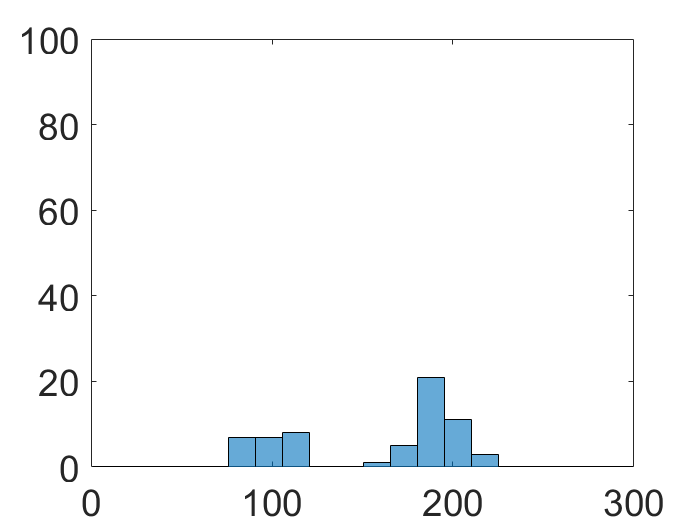}&
			\includegraphics[width=1.4in]{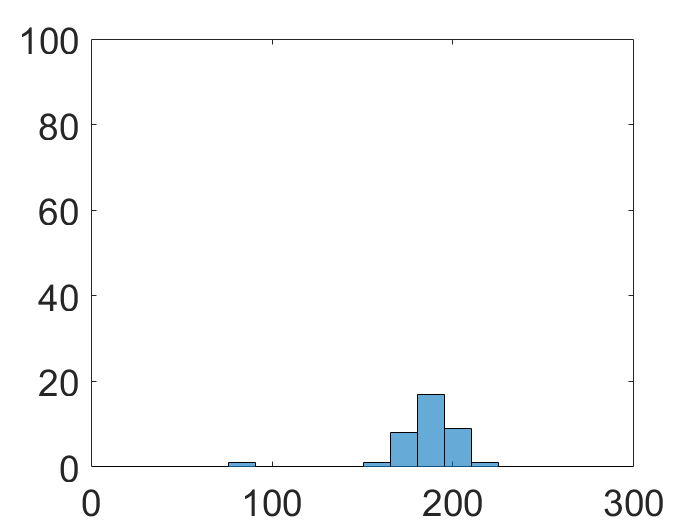} \\
			(Model 2.1, $n=4$) & (Model 2.1, $n=15$) &(Model 2.1, $n=50$) \\
			\includegraphics[width=1.4in]{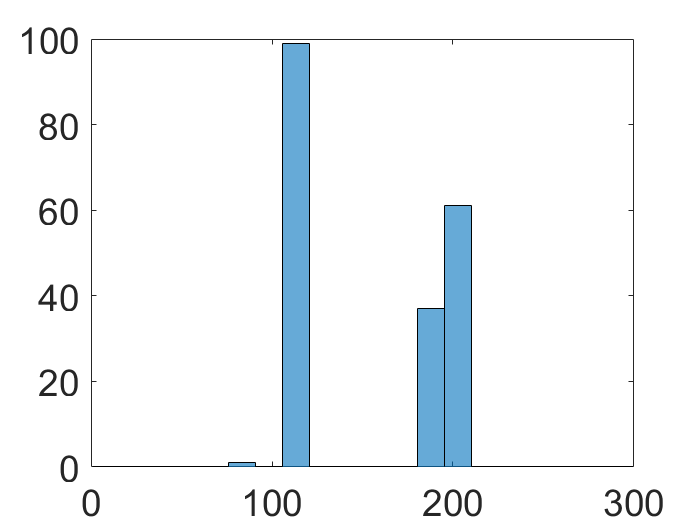}&
			\includegraphics[width=1.4in]{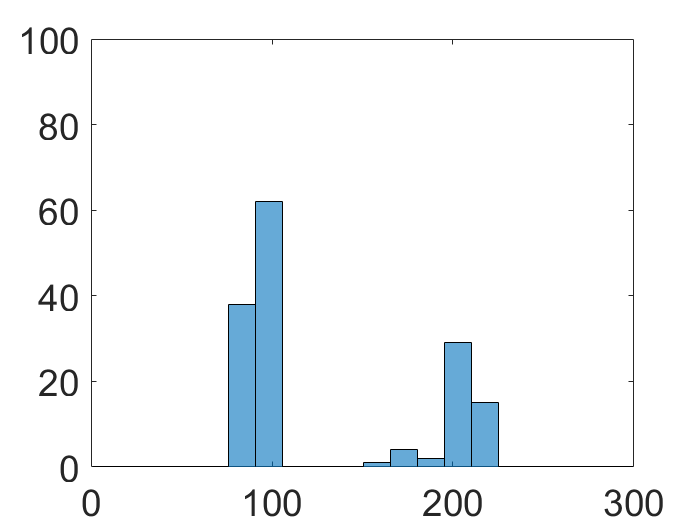}&
			\includegraphics[width=1.4in]{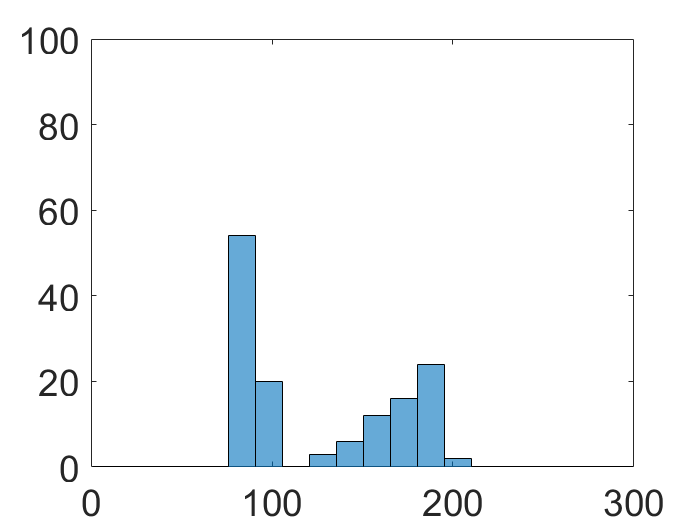} \\
			(Model 2.2, $n=4$) & (Model 2.2, $n=15$) &(Model 2.2, $n=50$) \\
			
		\end{tabular}
		\caption{ Results of {\bf Experiment 5}. Histograms of detected change-points given different trajectory models and dimensions with 100 runs in each case. Sample means and standard deviations for two change-points in Model 2.1 when n = 4, 15, 50 are (116.61, 15.57; 193.80, 12.63), (99.27, 12.24, 192.73, 12.31), (78, 0; 189.67, 11.85) respectively; Sample means and standard deviations for two change-points in Model 2.2 when n = 4, 15, 50 are (113.46, 4.19; 196.60, 5.30), (94.44, 7.32, 206.24, 13.14), (94.84, 17.14; 178.11, 12.65) respectively.} 
		\label{fig:sim4}
	\end{figure*}

	 In {\bf Experiments} $\bf 2-4$, a signal space series of length $300$ 
	 results in a covariance sequence of length $48$ (for $W=16$ and $S=6$). So, for consistency, we simulate covariance sequences of length 48 directly
	 in this case and apply our change-point detection algorithm with $100$ runs in each case. 
	 Finally, we map the detected change points back to the signal space time scale (with $300$ points) and display the histograms of detected change-points in Figure~\ref{fig:sim4}.
	 
	 In both the examples, the results show good performance in terms of locations and detection rates 
	 when $n$ is low or moderate. When $n$ is high the detection accuracy goes down, highlighting a limitation of 
	 the proposed method. As mentioned earlier, we are using a covariance-based representation and one requires
	 more time samples to keep the same level of precision in covariance estimation. 
	  Other factors, such as the nature of simulated time series, can also contribute to this lower performance.
	
	\iffalse
	To illustrate the algorithm in situations with known ground truth, 
	we simulate a $d$-dimensional time series data (with $900$ data points, i.e.  $T = 900$), 
	with three sub-intervals using three different multivariate normal distributions. 
	Data points $\{X_t\}$  from $t = 1$ to $t = 300$ are simulated from ${\cal N}(0,\Sigma_1)$, from $t=301$ to $t=600$ are  
	from  ${\cal N}(0,\Sigma_2)$, and from $t=601$ to $t=900$ are  from ${\cal N}(0,\Sigma_3)$, 
	where $\Sigma_1,\Sigma_2$ and $\Sigma_3$ are randomly generated in such a way that
	$d_{\tilde{P}}(\Sigma_i , \Sigma_{i+1})>0.6$. 
	To detect change-points in this time series,  we implemente Algorithm 2 using window parameters $W=30$, $S=20$, 
	where $W$ stands for the window size for calculating the sample covariance matrices and $S$ represents the step size of skipping for sliding windows.  
	Figure~\ref{fig:sim1} shows a plot of the test statistic, 
	$Z(\tau)$ versus $\tau$, for two cases,  $d =2$ in the left panel and  $d = 10$ in the right panel. 
	The detected change-points are marked using black circles, and one can 
	clearly see that the test statistics peak at time points that are around  
	the ground truth.
	
	\begin{figure}
		\begin{center}
			\begin{tabular}{cc}
				\includegraphics[width=2.5in]{figs/simulation,d=2.jpg}&
				\includegraphics[width=2.5in]{figs/simulation,d=10.jpg} \\
				$d=2$ & $d = 10$ 
			\end{tabular}
			\caption{Change-point detection on simulated data, with detections marked using circles.} \label{fig:sim1}
		\end{center}
	\end{figure}

    \fi

\subsection{Real Data: Single Subject Dynamic FC}

	Next we consider task fMRI (tfMRI) data from the HCP database. 
	The majority of the HCP tfMRI data were acquired at 3T, which is considered to be the field strength 
	currently most suitable for acquiring high quality data reliably from a large cohort of subjects. 
	Acquisitions are based on the blood oxygen level dependent (BOLD) contrast. 
	A series of 4D imaging data were acquired for each subject while they were performing tasks involving different neural systems, e.g. visual, motion or cognition systems. The acquired images
	have an isometric spatial resolution of $2$ mm and temporal resolution of $0.7$s. All fMRI data in HCP are preprocessed by removing spatial distortions, realigning volume to compensate for subject motion, registering the fMRI to the structural MRI, reducing the bias field, normalizing the 4D image to a global mean, masking the data with the final brain mask and aligning the brain to a standard space (\cite{Glasser2013}). This preprocessed tfMRI is now ready for FC analysis. 
	
	To map FC, we begin with the segmentation of  brain into regions using an existing template, 
	such as the  Automated Anatomical Labeling (AAL) atlas (\cite{aal2002}).  Time series for each region in tfMRI are extracted using CONN functional connectivity toolbox (\cite{pmid22642651}) and the default atlas in CONN toolbox. A [0.008, inf] (Hz) high-pass filter is applied to de-noise the time series data. We evaluate dynamic FC using the proposed method  for two different tasks: (1) gambling and (2) social cognition. The experimental results are categorized below. 
	{\bf We discuss a way of validating these results, later in Section 4.4. Here we present the experimental results 
	as a demonstration of our approach.}

\begin{enumerate}

\item {\bf Gambling Task}: 
		The gambling task in HCP was adapted from the one developed in \cite{pmid11110834}. Participants play a card guessing game in which 
		they are asked to guess the number on a mystery card in order to win or lose money.  Three different type of blocks are presented throughout the task: reward blocks, 
		neutral blocks and loss blocks. Brain regions that naturally relate to this gambling task include basal ganglia, ventral medial prefrontal and orbito-frontal. The basal ganglia contains multiple subcortical nuclei and is the part of brain that influences motivation and action selection. We selected four regions inside the basal ganglia: right Caudate, left Caudate, right Putamen and left Putamen, and studied FC using $4 \times 4$ SPDMs representing covariance of fMRI signal in these four regions.  Algorithm 2 was applied and the results are shown in Figure~\ref{fig:gamblingroi99_102} (a). It shows the evolution of 
		the test statistic $Z(\tau)$ versus $\tau$ for two different choices of window size $W = 16, 18$. 
		Despite different window sizes, the algorithm consistently detects the same two change-points, resulting in
		three subintervals where FC is statistically stationary. For each of these subintervals, we estimated the corresponding connectivity graphs using graphical Lasso and display them in  \ref{fig:gamblingroi99_102}(b).
		
		As another example under the same task -- gambling -- Figure~\ref{fig:networkroi10} shows results using 10 ROIs in the Temporal Lobe (including Heschl gyrus, superior temporal gyrus and middle temporal gyrus). This figure shows two detected change-points in (a) and reconstructed graphs for each sub-interval in (b). Compared with the four ROIs case in Figure~\ref{fig:gamblingroi99_102}, the change-points detected in Figure~\ref{fig:networkroi10} are much closer to each other in time. {\bf The detection results in these two 
		examples differ because the experiments involve different ROIs and subjects.}
		
		\begin{figure}
			\centering  
			
			\includegraphics[width=5.0in]{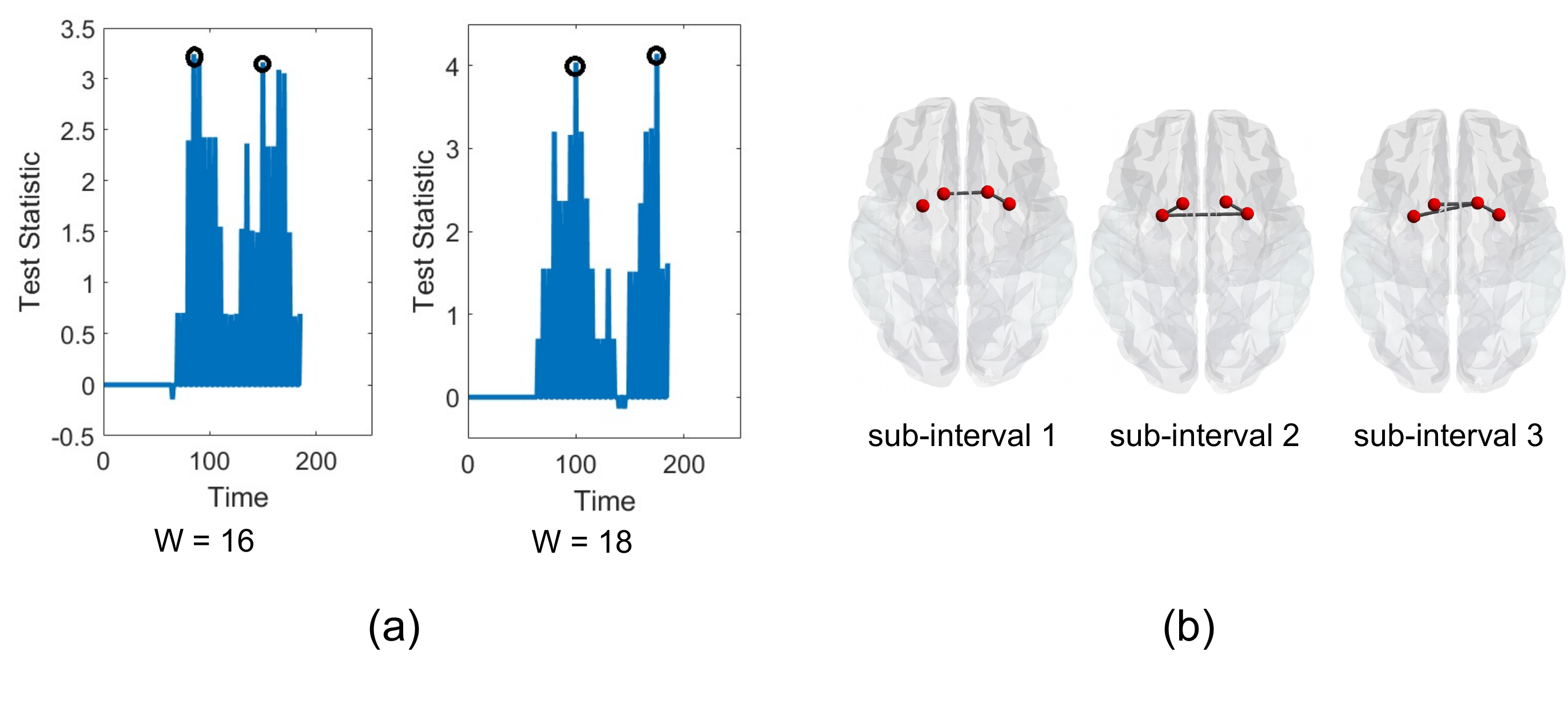}  
			\caption{Dynamic FC change-point detection for four regions in basal ganglia in gambling task. Peaks in change intervals in (a) are located at $t = 90, 150$ when $W=16$ and $t = 98, 176$ when $W=18$.} \label{fig:gamblingroi99_102}
		\end{figure}
		
		\begin{figure}
			\centering  
			\begin{tabular} {cc}
				\includegraphics[width=2.2in]{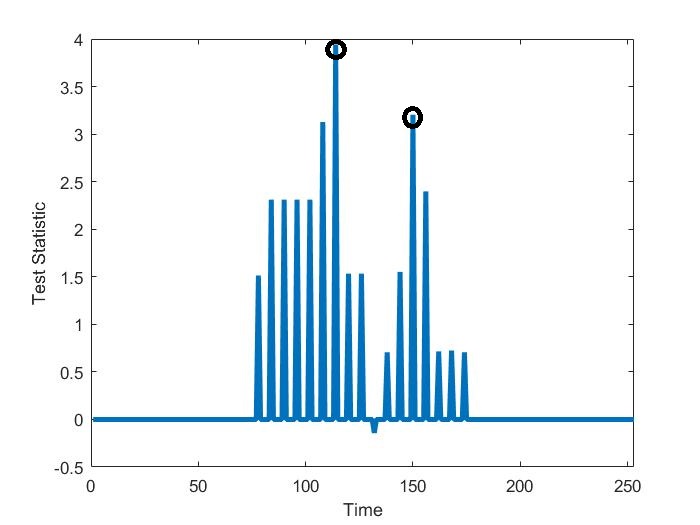}  &
				\includegraphics[width=2.8in]{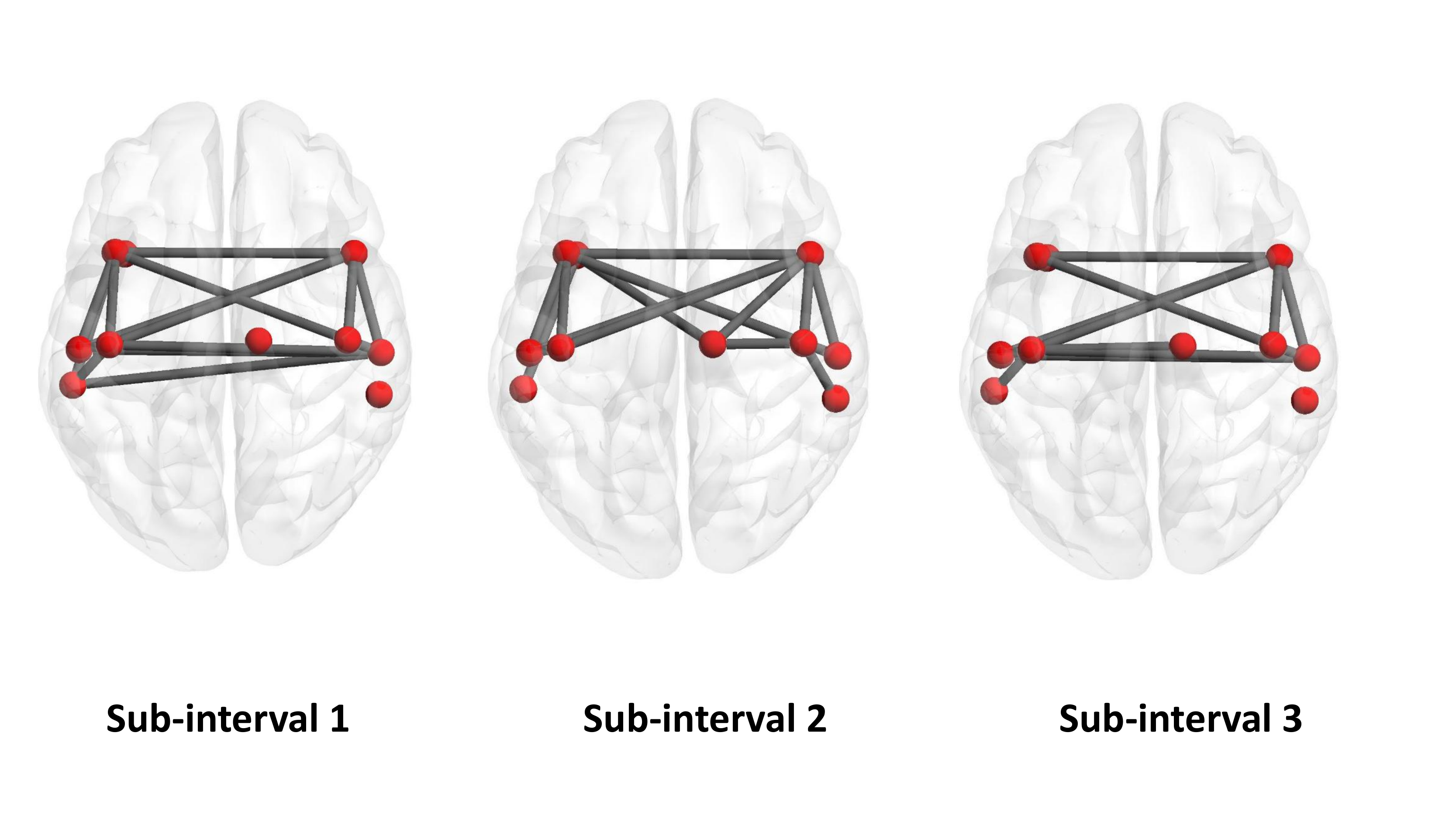} \\
				(a) & (b) \\
			\end{tabular}
			\caption{ FC change-point detection in gambling task for 10 regions in the Temporal Lobe  (ROI 78-87). (a) test statistic. Peaks in change intervals are located at $t =126, 198$. (b) reconstructed stationary FC networks for the sub-intervals.} 
			\label{fig:networkroi10}
		\end{figure}
		
		\item {\bf Social Cognition Task:}
		In this task the
		participants were presented with short video clips of objects (squares, circles, triangles) that either 
		interact in some way, or moved randomly on the screen (\cite{Castelli2000314}). After each video clip, participants judge whether the objects had  (i) a mental interaction (an interaction that appears as if the shapes are affecting 
		feelings and thoughts); (ii) not sure or (iii) no interaction. Each run of the task has 5 video blocks (2 Mental and 3 Random in the data used
		here). Possible brain regions that relate to this task are medial prefrontal cortex, temporal parietal junction, inferior and superior temporal sulcus (\cite{Barch2013}). 
		Since this task has more time blocks (5 blocks) than the gambling data, 
		we expect to detect more change-points.  
		In this example, we used $14$ regions taken from Occipital lobe, Parietal lobe and Temporal lobe, and the algorithm 
		detects several change-points. Figure~\ref{fig:networkroi14} (a) shows the result using { $W = 16$}. We detected three major change-points 
	        and reconstructed the connectivity graphs for these $14$ ROIs within the four sub-intervals of stationary FC. 
		Figure~\ref{fig:networkroi14} (b) shows the reconstructed networks. We note that the intermediate 
		networks (during intervals 2 and 3) are more complex than the networks at the two ends. 
		
		%Another example with the social cognition task data is shown in Figure. \ref{fig:networkroi8} using 8 ROIs. After detecting two change-points as shown in (a), we reconstructed the connectivity in each sub-interval as in (b). 
		
		%\begin{figure}
		%	\centering  
		%	\begin{tabular} {cc}
		%		\includegraphics[width=2.5in]{figs/AAL_Social_Z_45-52.jpg}  &
		%		\includegraphics[width=3.0in]{figs/Social_45-52_AllStages.pdf} \\
		%		(a) & (b) \\
		%	\end{tabular}
		%	\caption{FC change-point detection in social task for 8 regions (ROI 45-52). (a) test statistic. (b) reconstructed FC networks for the sub-intervals.} 
		%	\label{fig:networkroi8}
		% \end{figure}
		
		\begin{figure}
			\centering  
			\begin{tabular} {cc}
				\includegraphics[width=2.2in]{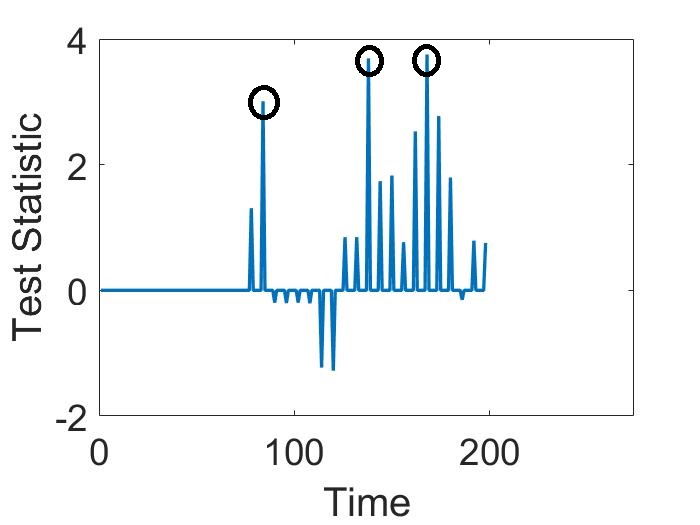}  &
				\includegraphics[width=2.8in]{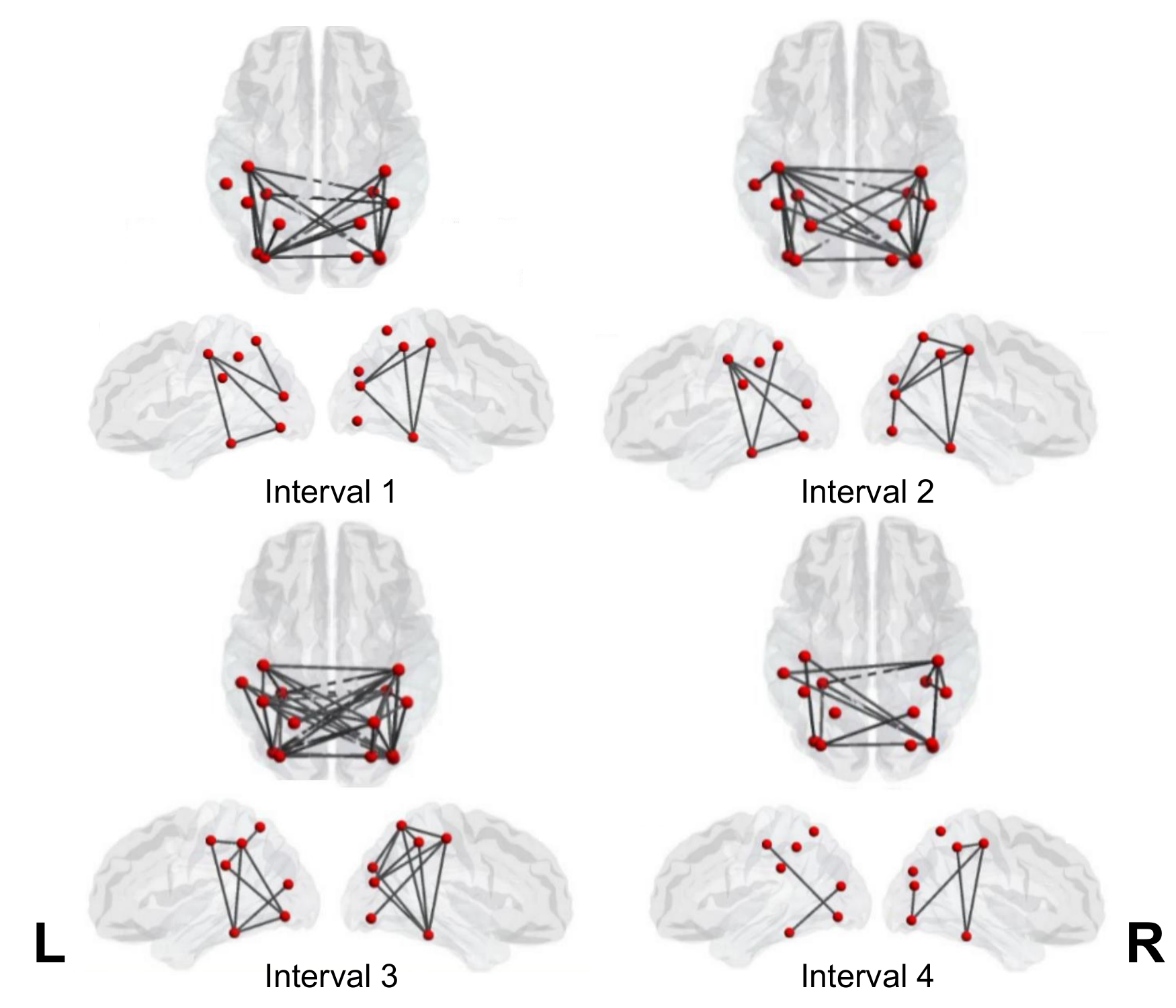} \\
				(a) & (b) \\
			\end{tabular}
			\caption{ FC change-point detection in social cognition task for 14 regions (ROI 50-63). (a) change-points with $W = 16$. Peaks in change intervals are located at $t =84, 138, 168$. (b) reconstructed FC networks for the four intervals.} 
			\label{fig:networkroi14}
		\end{figure}
	\end{enumerate}

\subsection{Real Data: Dynamical FC for Multiple Subjects}
		Next we use tfMRI data from gambling task and social cognition task to study change patterns across a large set of subjects (197 subjects) from the HCP.
		As in the earlier experiments, we begin with the segmentation of brain into multiple regions using an existing template as in \cite{Gordon2016}. 
		The cortex area is divided into a few networks as visual, auditory, default, frontoparietal, dorsal attention, cingulo-opercular, ventral attention, and salience network. Each network contains several ROIs.
		One fMRI time series for each selected region is extracted using CONN functional connectivity toolbox (\cite{pmid22642651}). In the following experiments, we used a window size of
		$W= 16$ and a moving step size of $S=6$ (as in previous experiments).
		
		Since it is hard to estimate locations of change-points precisely due to: 
		(1)  sliding windows and autocorrelation, and (2) 
		temporal misalignment between subjects,  and (3) 
		different response times of individuals to the stimuli in the tasks, we expect that different subjects will have 
		different change-point patterns. The methodology used for change-point detection can also potentially introduce a 
		temporal shift in the detected points. A direct comparison of the raw change-points turns out to be ineffective. After detecting the change-points, we applied Algorithm \ref{algo:alignment} to reduce the temporal misalignment and to discover the common change-point patterns across subjects. Note that in the following experiments (
		labeled as Setup 1-6), we use the same 197 subjects and use the same layout of pictures to present the results 
		(Figure~\ref{fig:CPCingulerOpercGambling}-\ref{fig:CPVentralRest}).  Taking Figure~\ref{fig:CPCingulerOpercGambling} as an illustrative example, in panel (a) and (b), we show the raw and aligned detected change-points across 197 subjects.  In these panels, each colored line represents the change-point location for an individual subject.  Panel (c) shows  the smooth functions $\{f_i\}$ 
		(fitted to the test statistic curves $\{Z_i\}$)
		before the alignment and panel (d) shows the aligned functions $\{\widetilde{f_i}\}$. The corresponding warping functions are displayed in panel (e). The specific ROI names for each figure are given in the supplementary material. We present extensive results using the following 
		experimental setup.

\begin{itemize} 

\item Setup 1: 12 ROIs are chosen from the CinguloOperc network during the gambling task. The results are shown in Figure~\ref{fig:CPCingulerOpercGambling}.
	
\item Setup 2: 	
	A different set of 14 ROIs are picked from the DorsalAttn network during gambling task. The results are  displayed in Figure~\ref{fig:CPDorsalGambling}. 
	
\item Setup 3: 	
	12 ROIs (same ROIs as in Setup 1) are used from the CinguloOperc network during the social cognition task. The results are shown in Figure~\ref{fig:CPCingulerOpercSocial}. 
	
\item Setup 4: 	  
    14 ROIs are picked from the DorsalAttn (same ROIs as in Setup 2) network during the social cognition task.
	Figure~\ref{fig:CPDorsalSocial} shows the results. 
	
\item Setup 5: 	
	10 ROIs are picked from the DorsalAttn network during the resting state. The results are shown in Figure~\ref{fig:CPDorsalRest}.
	
\item Setup 6: 	A different set of 10 ROIs from the VentralAttn network are picked for the resting state data. The results are shown in Figure~\ref{fig:CPVentralRest}.

	\begin{figure}
		\centering  
		\includegraphics[width=5in]{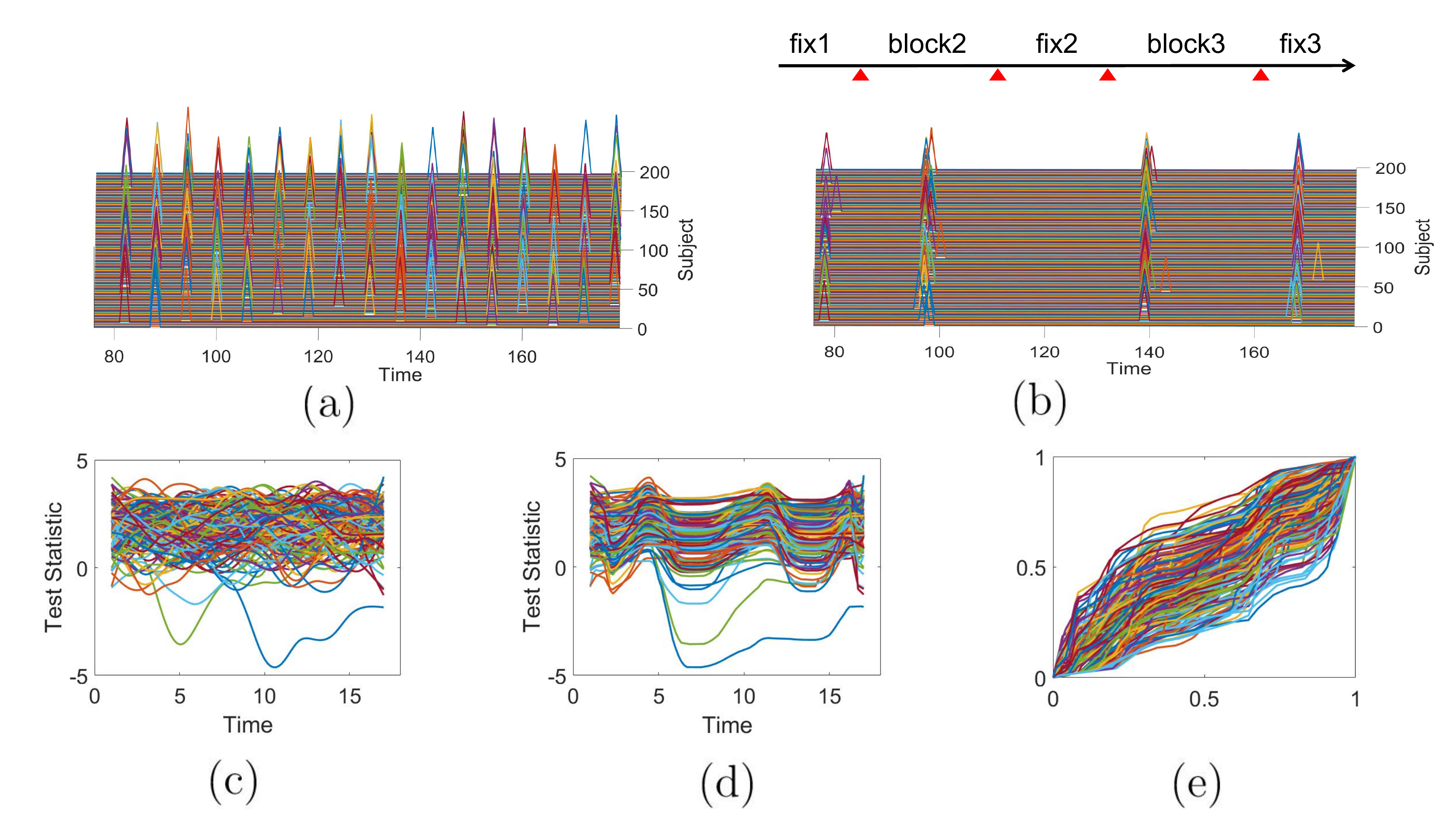}  
		\caption{Change-point patterns of 12 ROIs in the CingulerOperc network during the gambling task for 197 subjects. (a) shows $\{C_i\}$ of the detected change-points before alignment, (b) shows $\{\widetilde{C_i}\}$ after the alignment, (c) shows the fitted smooth test score functions $\{f_i\}$, (d) shows the aligned test score functions $\{\widetilde{f_i}\}$ and (e) shows the warping functions for alignment.} 
		\label{fig:CPCingulerOpercGambling}
	\end{figure}	
	
	%2. Dorsal Attention gamling task: Figure.\ref{fig:CPDorsalGambling}
	%NewparcelGambling200,116,158,192,202,206,211,214,239,253,255-256,265,274,278
	\begin{figure}
		\centering  
		\includegraphics[width=5in]{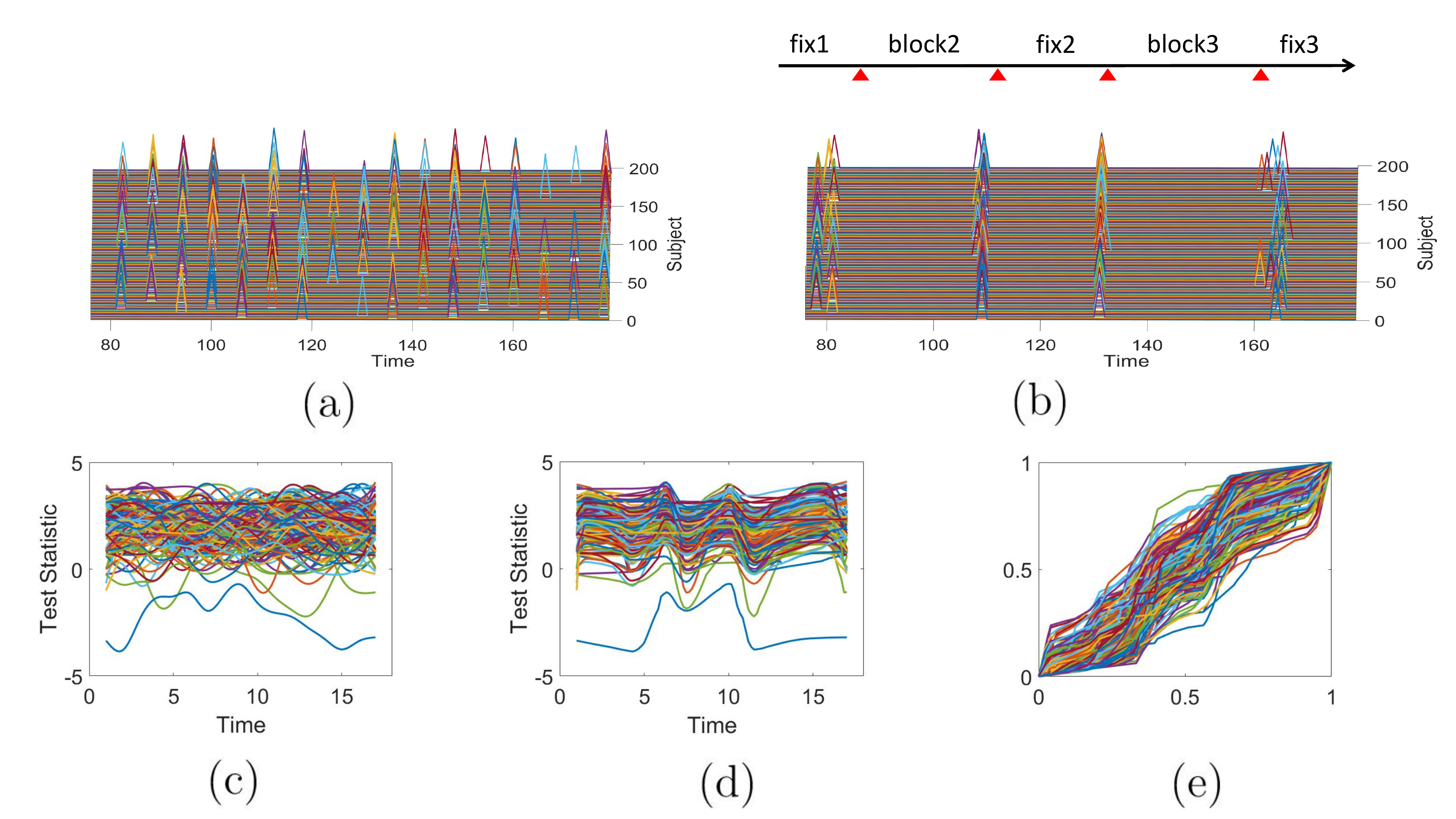}  
		\caption{Change-point patterns of 14 ROIs in the DorsalAttn network during the gambling task for 197 subjects. }
		\label{fig:CPDorsalGambling}
	\end{figure}	
	
	%3. CinguloOPerc social task: Fig.\ref{fig:CPCingulerOpercSocial}
	%NewparcelSocial200,222,226,237,238,241,248,249,251,252,277,320,321
	\begin{figure}
		\centering  
		\includegraphics[width=5in]{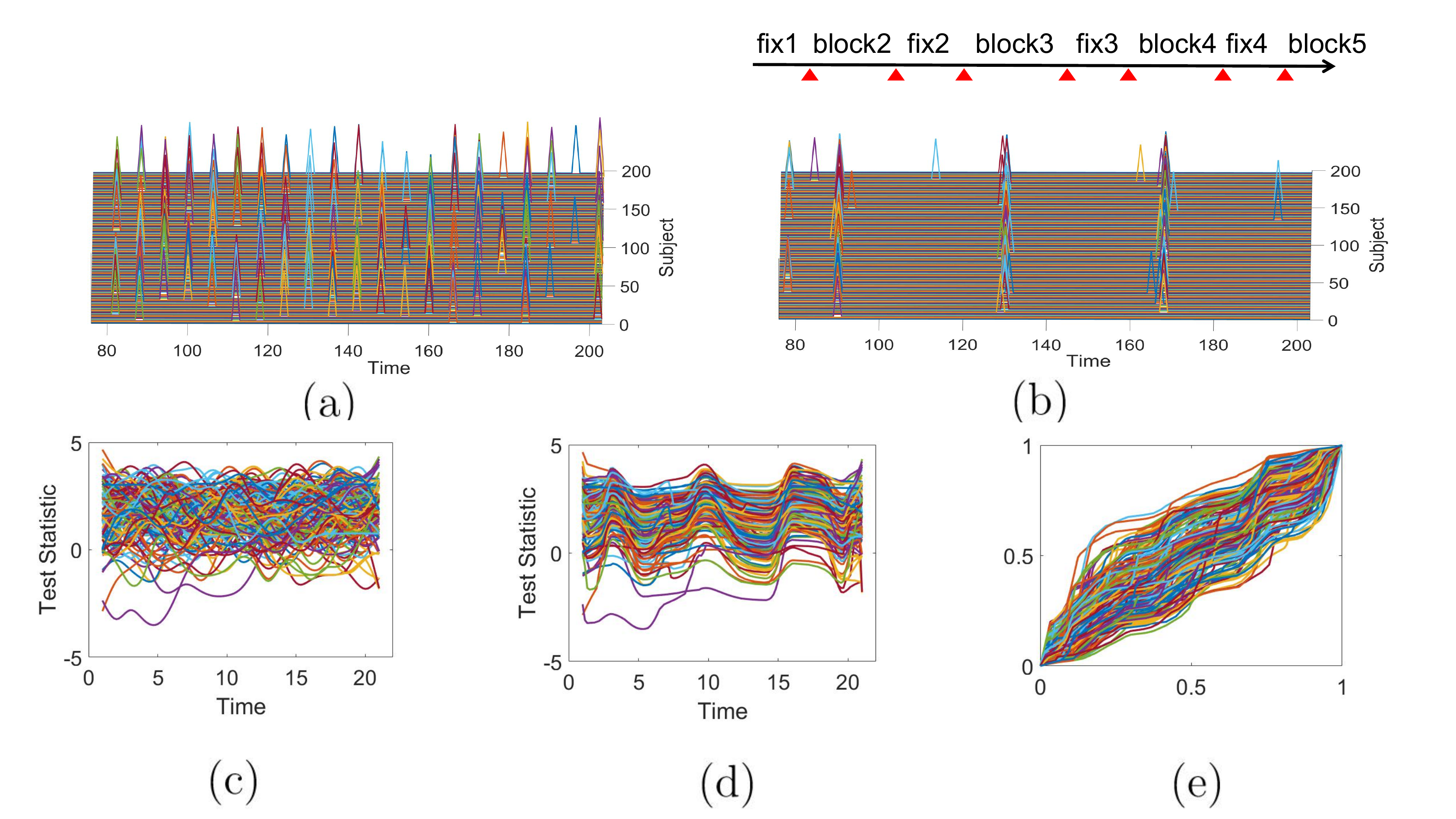}  
		\caption{Change-point patterns of 12 ROIs in the CinguloOperc network during the social task for 197 subjects.} 
		\label{fig:CPCingulerOpercSocial}
	\end{figure}		
	
	%4. Dorsal Attention social task: Fig.\ref{fig:CPDorsalSocial}
	%NewparcelSocial200,116,158,192,202,206,211,214,239,253,255-256,265,274,278
	\begin{figure}
		\centering  
		\includegraphics[width=5in]{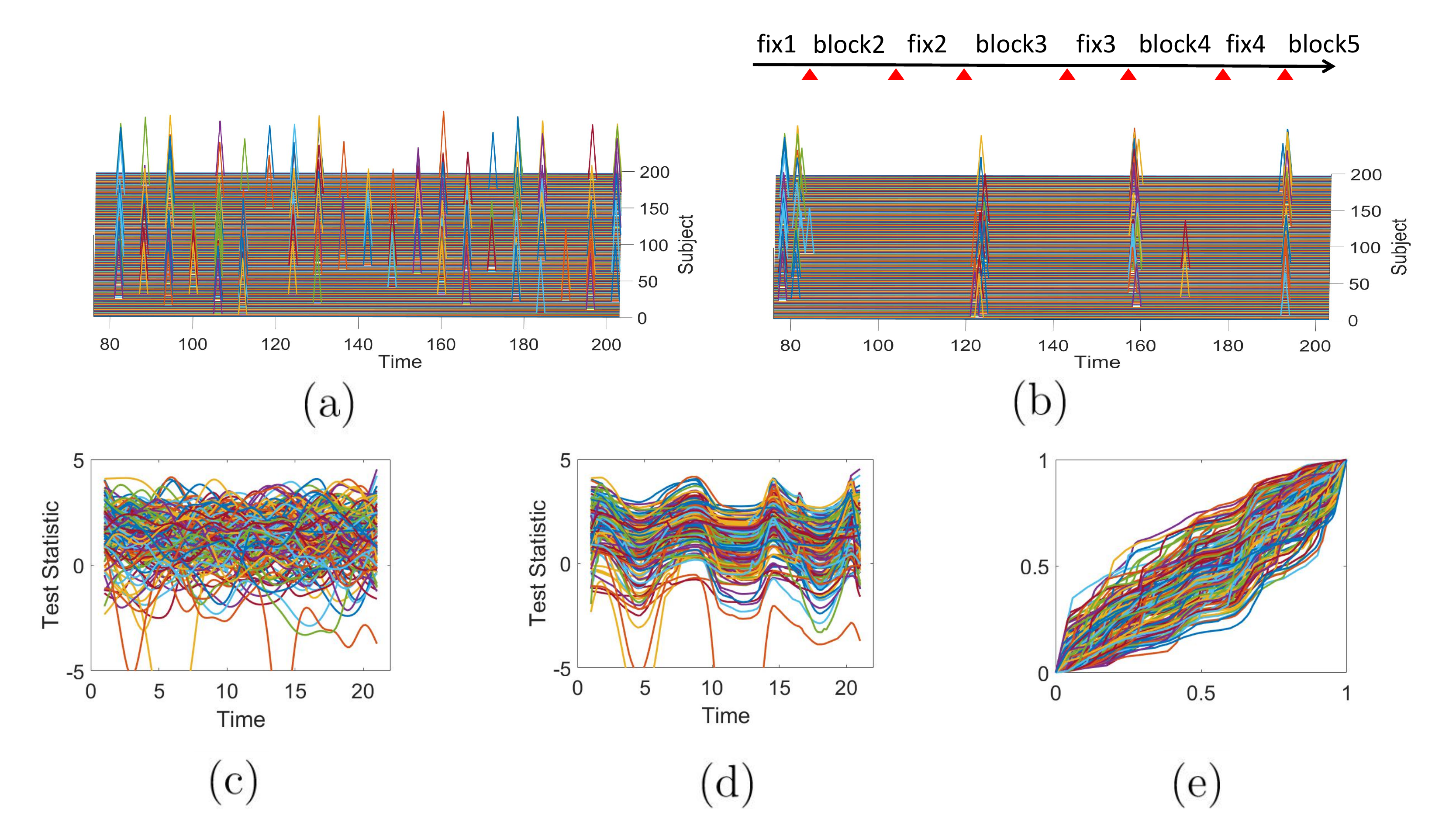}  
		\caption{Change-point patterns of 14 ROIs in the DorsalAttn network during the social task for 197 subjects.} 
		\label{fig:CPDorsalSocial}
	\end{figure}		
	
	%5. Dorsal Attention resting state: Fig.\ref{fig:CPDorsalRest}
	%NewparcelRest200,206,211,214,239,253,255,256,265,274,278
	\begin{figure}
		\centering  
		\includegraphics[width=5in]{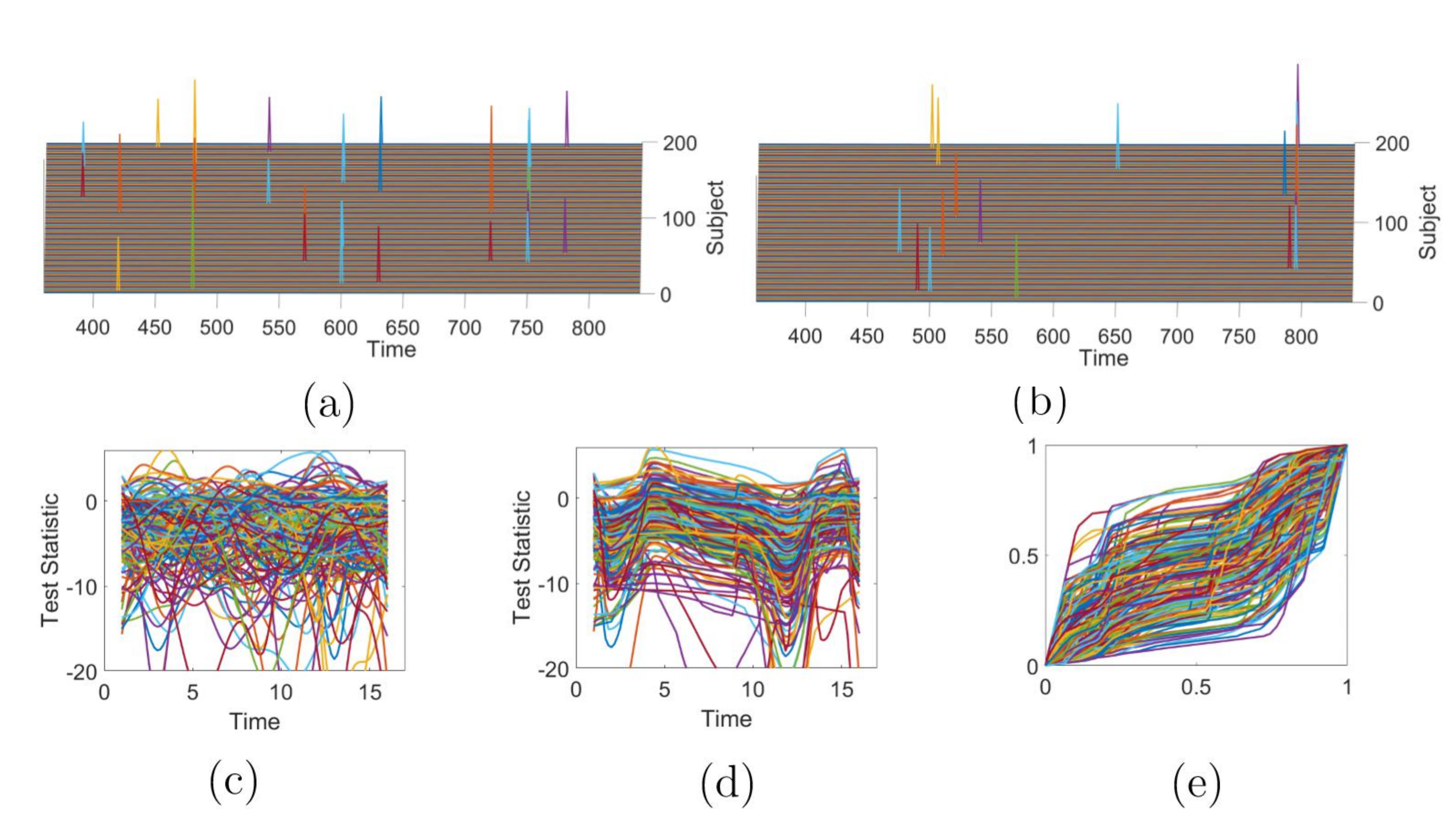}  
		\caption{Change-point patterns of 10 ROIs in the DorsalAttn network during the resting state for 197 subjects.} 
		\label{fig:CPDorsalRest}
	\end{figure}	
	
	%6. Ventral Attention resting state: Fig.\ref{fig:CPVentralRest}
	%NewparcelRest200,26,63-65,82-83,161,164,224-225
		\begin{figure}
		\centering  
		\includegraphics[width=5in]{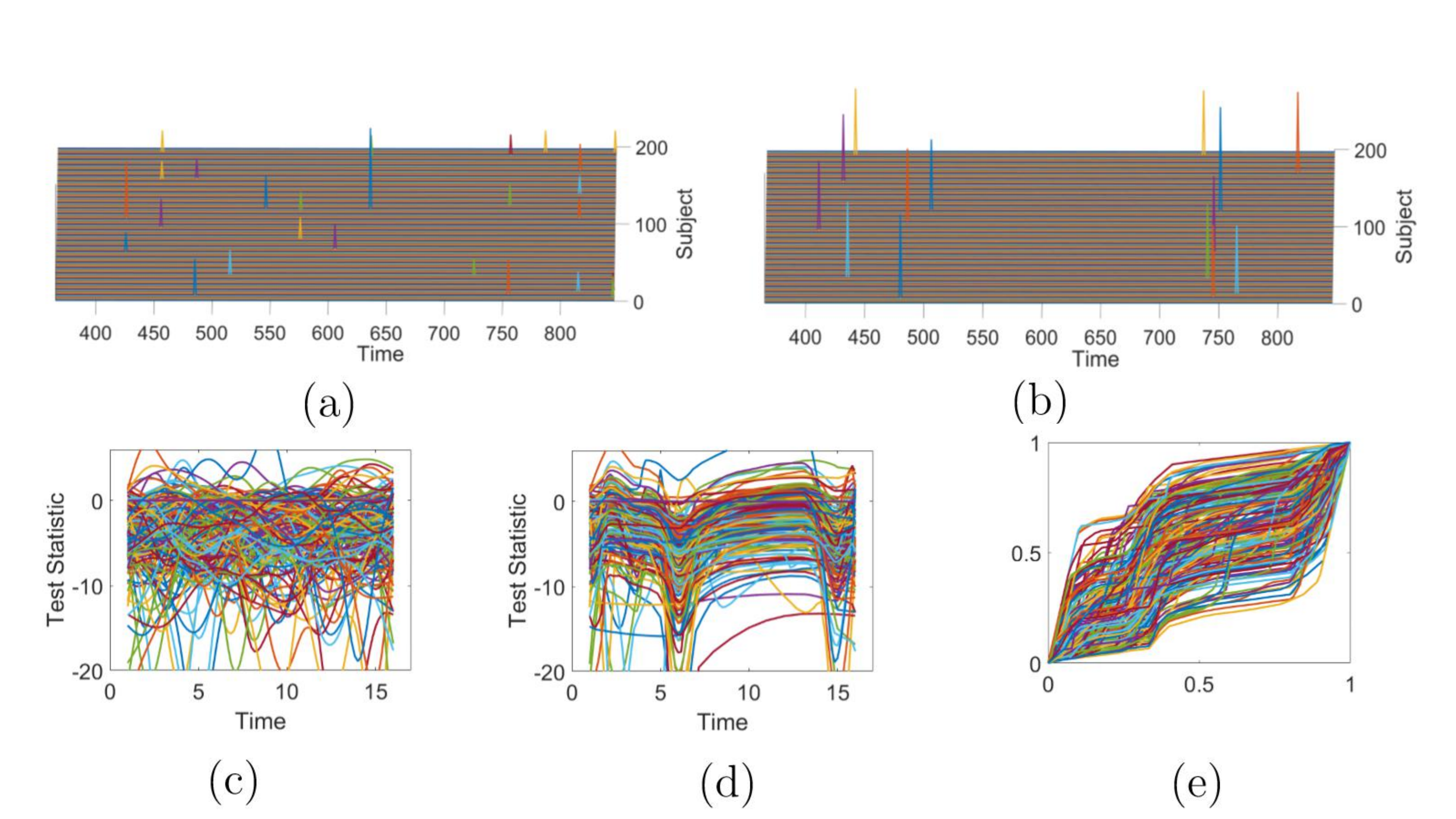}  
		\caption{Change-point patterns of 10 ROIs in the VentralAttn network during the resting state for 197 subjects.} 
		\label{fig:CPVentralRest}
	\end{figure}	

\iffalse
	\begin{figure*} 
		\centering
		\begin{tabular}{ccc}
			\includegraphics[width=1.8in]{figs/NewparcelRest200,206,211,214,239,253,255,256,265,274,278,Zpeaks,beforealignment.png}&
			\includegraphics[width=1.8in]{figs/NewparcelRest200,206,211,214,239,253,255,256,265,274,278,Zpeaks,afteralignment.png}\\
			(a) & (b) \\		
			\includegraphics[width=1.5in]{figs/NewparcelRest200,206,211,214,239,253,255,256,265,274,278,Z,beforealignment.png}&
			\includegraphics[width=1.5in]{figs/NewparcelRest200,206,211,214,239,253,255,256,265,274,278,Z,afteralignment.png}&
			\includegraphics[width=1.5in]{figs/NewparcelRest200,206,211,214,239,253,255,256,265,274,278,gamma.png} \\
			(c) & (d) &(e) \\						
		\end{tabular}
		\caption{Change-point patterns of 10 ROIs in the Ventralattn network during the resting state for 197 subjects. } 
		\label{fig:CPVentralRest2}
	\end{figure*}
\fi

\end{itemize}

    As one can see that the time warping involved in alignment of these datasets is quite significant. While it is difficult to discern any pattern in the original detections, 
    there are certainly common pattens in numbers and locations of change-points across subjects after the alignment in all of these cases. 

\begin{figure}
	 \centering  
	  \begin{tabular} {cc}
		\includegraphics[width=2.5in]{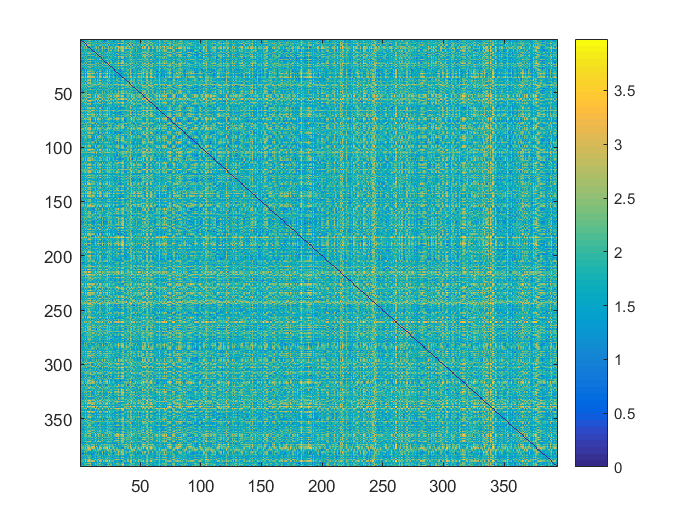}  &
		\includegraphics[width=2.5in]{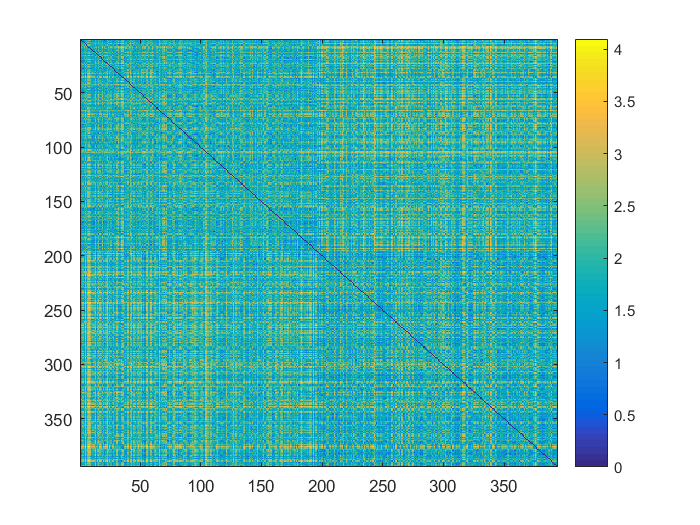} \\
		(a) & (b) \\
   	  \end{tabular}
	 \caption{ Pairwise distances between 394 test statistic functions from both gambling task (first 197 functions) and social task (second 197 functions) given same 12 ROIs in CinguloOperc network. (a) pairwise distances between  $f_{i}s$; (b) pairwise distances between $\widetilde{f_i}s$.} 
	 \label{fig:Dist_NotAligned_CinguloOperc_GS}
    \end{figure}

	In order to emphasize the effect of alignment further, we compute and display a matrix of pairwise distances between 
	the test statistic functions of different subjects. In this experiment we take pairwise distances between 394 test statistic functions from both gambling task (first 197 functions) and social task (second 197 functions) for the same 12 ROIs in CinguloOperc network as in Setup 1 and Setup 3.  Figure~\ref{fig:Dist_NotAligned_CinguloOperc_GS} (a) shows the matrix of distances between fitted test statistic functions before alignment ($\{f_i\}$), and Figure~\ref{fig:Dist_NotAligned_CinguloOperc_GS} (b) shows the pairwise distance matrix between fitted test statistic functions after alignment ($\{\widetilde{f_i}\}$), using Algorithm \ref{algo:alignment}.
	In Figure~\ref{fig:Dist_NotAligned_CinguloOperc_GS} (b) we observe a stronger diagonal block pattern than in Fig. \ref{fig:Dist_NotAligned_CinguloOperc_GS} (a), which shows the effect of temporal alignment for task classifications.

	\subsection{Pattern Validation Using Experiment Design Structures}
 A large number of experiments performed using different tasks and  different ROIs combinations reveal common patterns in change-points in FC across subjects. 
 However,  an important question is: How can one validate these results in the 
 absence of a ground truth?  In this section we take a careful look at the experiment designs and try to compare them with the detected patterns. The 
 overlapping nature of designs blocks and the change-point patterns provide a certain validity to the conclusions. 
	 
   Let us start with Setup 1 and Setup 2 that are related to the gambling task. 
   The gambling task is presented in blocks of 8 trials that are either mostly reward or mostly loss blocks.  In each of the two runs in HCP data acquisition, there are 2 mostly reward and 2 mostly loss blocks, interleaved with 4 fixation blocks (15 seconds each). In Fig. \ref{fig:CPCingulerOpercGambling} (b), the timeline drawn at the top of the picture gives an example of gambling task blocks taking place, where ``fix1'' represents the first fixation block, and ``block2'' represents the second task block (either mostly reward or mostly loss), divided by small red triangles. We can see the aligned change-points mostly take place at intervals between task blocks and fixation blocks. The same holds for the results shown in Fig.  \ref{fig:CPDorsalGambling} (b). In other words, 
   the aligned change-points correspond to the design of the experiment used in collected tfMRI data. 
	
	In social task experiments (Setup 3 and Setup 4), our change-point detection model is implemented on the same ROIs as in the gambling task. For this task the participants were presented with short video clips (20 seconds). Each of the two task runs has 5 video blocks and 5 fixation blocks (15 seconds each). 
	In this case we expect to detect more change-points between task or fixation blocks than in gambling task experiments. Comparing Fig. \ref{fig:CPCingulerOpercSocial} and Fig. \ref{fig:CPDorsalSocial}, we see although the ROIs are the same as in gambling task experiments, the change-point patterns are quite different. Furthermore, the aligned change-points mostly lie in time intervals between task blocks and fixation blocks, given the task timeline at the top of pictures in Fig. \ref{fig:CPCingulerOpercSocial}(b) and Fig. \ref{fig:CPDorsalSocial}(b).
	
	In the remaining two experiments (Setup 5 and Setup 6) we consider the resting state data. These
	rfMRI data are acquired in four runs of approximately 15 minutes each, two runs in one session and two in another session, with eyes open with relaxed fixation on a projected bright cross-hair on a dark background (and presented in a darkened room). 
	In Fig. \ref{fig:CPDorsalRest} and Fig. \ref{fig:CPVentralRest} we observe very few change-points at random time points either before alignment (in (a))
	or after alignment (in (b)).

    The general correspondence between aligned change-point patterns and experiment block designs, for both gambling and social tasks, 
    can be seen as a validation for our results. The results for resting state experiments where we observe only few change-points also meet our expectation.
	
	\section{Conclusion and Discussion}
	In this paper we have studied patterns of changes in dynamic FC during task-related stimuli across human populations. The methodology consists of a pipeline of 
	techniques used to extract and analyze changes in FC over time for individual subjects. We represent short-term FC as a covariance matrix and model its 
	temporal evolution as a time-series on the manifold of SPDM matrices. Using a Riemannian structure on this manifold, and a graphical approach involving minimal 
	spanning tree, we detect change-points in FC for individual subjects under different tasks and different ROIs. When comparing these detections across subjects, it is difficult to
	discern common patterns due to the variability introduced by proposed method and misalignment of responses by different people. Using a simple technique for time-warping based alignment of change-point statistics, we are 
	able to visualize similarity of change-point patterns across a large sample of subjects. In order to validate these results, we find general correspondence between these change-point patterns and the 
	block designs of experiments used in the HCP tfMRI data collection.
	
	Our proposed method naturally has some limitations and we list some of them here. 
	One limitation is that the theory in \cite{chen2015} requires {\it i.i.d.} samples, while we
	have applied it to a correlated time series. One should ideally 
	develop a change-point detection approach suited for a time series with dependent samples on a nonlinear manifold. 
	Another limitation is that our method is not suitable for detecting change-points at end points of a time series. This is
	not a major limitation since the change points are usually
	meaningful only away from the boundaries.	
	Lastly, since there are several user-specified algorithmic parameters that potentially influence the final result, there may be some variability
	in the detection results due to these choices. We have used extensive experiments with simulated data and known ground truth to emphasize the
	success of this approach but we acknowledge this dependence of results on algorithmic choices. 
	
	In future, we would like to involve additional datasets in such studies, in order to detect and differentiate patterns from different subpopulations. In order to improve
	overall performance, we would like to investigate improvements in different pieces of the proposed pipeline. 
	For instance, one can replace the current MST based change-point detection procedure by a method that takes into account the 
	correlated nature of the SPDM time series, as mentioned above. 
	
	\section*{Acknowledgements}
	
	This research was supported in part by the NSF grants to AS --  NSF DMS CDS\&E 1621787 and NSF CCF 1617397. ZZ was supported in part by 
	SAMSI via grant NSF DMS 1127914 during this research. 
	
	The authors are thankful to the organizers of HCP database for making it available to public for this research. HCP is funded in part by 
	WU-Minnesota consortium (PIs: Van Essen and Ugurbil) 1U54MH091657. 
	
	\newpage
	\section{References}
	\bibliographystyle{model2-names}
	\bibliography{FCCPD}
\end{document}